\documentstyle[aps,prb,epsf]{revtex}
\begin{document}
\title{Lattice Pseudospin Model for $\nu=1$ Quantum Hall Bilayers}
\author{A.A. Burkov and A.H. MacDonald}
\address{\small Department of Physics, Indiana University, Bloomington, 
IN 47405}
\address{\small Department of Physics, The University of Texas at Austin, 
Austin, TX 78712} 
\date{\today}
\maketitle
\begin{abstract}

We present a new theoretical approach to the study of $\nu=1$ 
quantum Hall bilayer that is based on a systematic mapping of
the microscopic Hamiltonian to an anisotropic SU(4) spin 
model on a lattice.  To study the properties of 
this model we generalize the Heisenberg model Schwinger boson mean 
field theory (SBMFT) of Arovas and Auerbach to 
spin models with anisotropy.  We calculate the temperature 
dependence of experimentally observable quantities, including 
the spin magnetization, and the differential interlayer capacitance.
Our theory represents a substantial improvement over the conventional 
Hartree-Fock picture which neglects quantum and thermal fluctuations, 
and has advantages over long-wavelength effective models that 
fail to capture important microscopic physics at all realistic layer 
separations. The formalism we develop can be generalized to treat 
quantum Hall bilayers at filling factor $\nu=2$.  

\end{abstract}
\newpage
\section{Introduction}

The physics of high-mobility two-dimensional electron systems in strong 
perpendicular 
magnetic fields continues to produce surprises.  Electronic states 
with strong correlations in this regime originate fundamentally from the 
macroscopic 
degeneracies, Landau levels, in the spectrum of the free-particle kinetic
energy Hamiltonian.  When spin, layer, or other degrees of freedom are 
important at low energies, correlations at integer filling factors 
can give rise to broken symmetry ground states~\cite{Jungwirth00} with 
interesting physical properties.  
Bilayer quantum Hall systems, whose ground states can have spontaneous 
interlayer 
phase coherence, provide the simplest and best characterized example 
of this tendency, and have been studied both theoretically and experimentally 
for 
more than ten years.~\cite{Fertig89,Macdonald90,Brey90,Wen92,Shayegan,Iwazaki,Eisenstein94,Moon95,Yang96,Zheng97,Dassarma98,Ezawa,Dassarma99,Macdonald99,Schliemann00,Schliemann01,Burkov02,Rajaraman00,Radzihovsky01,Ramin01,Eisensteinrefs,josephson} 
The superfluid effects associated with spontaneous coherence are especially
dramatic.~\cite{Eisensteinrefs}  
Most work on these states has concentrated on the case of total lowest
Landau level (LLL) filling factor $\nu$ equal to one.
The quantum Hall effect, indicative of a gap for charged excitations, 
is observed in this system at sufficently small interlayer
separations $d$.  The charge gap in this case must result entirely from
interlayer correlations,~\cite{Halperin93} since individual 
layer states at filling factor $\nu=1/2$ are compressible.
As the layer separation or the electron density in each layer is decreased
there is a phase transition, likely of first order in the absence of disorder,
from a compressible state with no quantum Hall effect
to a highly correlated incompressible 
state~\cite{Macdonald90,Eisenstein94,Schliemann01} with spontaneous coherence.
In this paper we present a new approach which can be used to study the 
properties 
of these interesting ordered states and how they vary with layer separation. 

The layer degree of freedom in the quantum Hall bilayer can be conveniently
described using a pseudospin language: pseudospin up corresponds to an 
electron in the top layer and pseudospin down to an electron in the 
bottom layer.~\cite{Moon95}
The interlayer phase coherent quantum Hall state corresponds in this language 
to a pseudospin ferromagnet with an $\hat x - \hat y$ easy-plane.
Our work is motivated by two limitations that apply to most theoretical work
on bilayer quantum Hall ferromagnets. 
The usual assumption in most previous studies of $\nu=1$ bilayers
has been that the spin degree of freedom can be ignored, {\it i.e.} that
spin dynamics are frozen out by the magnetic field.
At finite temperatures, however, this assumption is usually not justified:
the typical value of the Zeeman gap is only $\sim$ 1K, which is often 
smaller than the pseudospin gap when the interlayer phase is fixed by 
finite tunneling.  There is a broad range of nonzero temperatures for many
samples in which there is a nontrivial interplay between spin and pseudospin
fluctuations.  
A second set of limitations that applies to much theoretical work on 
quantum Hall bilayers follows from the use of a gradient expansion in deriving
an effective continuum pseudospin model.  
It is sometimes bothersome in applying  
this approach that divergent coefficients, due to the long-range of the 
Coulomb interaction,
appear in the energy functional when one goes beyond 
the gradient expansion's leading order.  
A more severe limitation, however, is the 
fact that this approach cannot address the physics that gives rise to the 
phase transition between ordered and disordered states which, as we explain in
some detail, is due to competing interactions on microscopic length scales. 
The importance of correlations on microscopic length scales for
the phase transition was already evident in early work where it was associated
with soft collective excitations\cite{Macdonald90} at a finite wavevector.
The precise nature of the compressible state close to the transition is 
not completely understood at present and we belive that our approach can shed 
some light on this question. 
   
In this paper we present a careful study of the spin-pseudospin 
physics of the $\nu=1$ bilayer in the quantum Hall state which is free from 
these limitations.  
Our approach is based on a systematic mapping from the microscopic
interacting electron problem to a model with SU(4) generalized spins on 
a lattice and anisotropic interactions.
SU(4) group appears here since we take into account
both spin and pseudospin degrees of freedom and thus single-electron
states are four-component spinors.
The {\it only} simplification necessary to derive this mapping 
is the assumption of the absence of charge fluctuations,
reasonably well justified in the incompressible state.  
To take advantage of the existence of a charge gap in the quantum Hall regime, 
we use a complete orthonormal set of magnetic Wannier functions on a 
von Neumann lattice as the LLL orbital basis. 
Given this basis, and the neglect of charge fluctuations,
we can derive a coupled spin-pseudospin Hamiltonian which turns out  
to be unexpectedly simple.  Using the spin-Hamiltonian we derive, the 
full arsenal of analytic and numerical techniques that have been developed
to solve spin problems on a lattice can be applied to quantum Hall systems. 
In this paper, we present a study of this Hamiltonian using the 
Schwinger boson mean field theory,~\cite{Arovas88} developed originally 
by Arovas and Auerbach for the SU(2) invariant Heisenberg model, 
appropriately
generalized for the anisotropic case.  We obtain a variety of new results for 
the temperature dependence of some experimental observables, the spin 
magnetization 
and the interlayer differential capacitance in particular.  
We note that our approach is valid not only for the $\nu=1$ 
case but can be easily generalized to a quantum Hall bilayer with any 
integer filling factor, and with any orbital character of nearby Landau levels.
For bilayers, the case of $\nu=2$ is also interesting and 
has been studied extensively.~\cite{Zheng97,Dassarma98,Dassarma99,Macdonald99,Schliemann00}
Since the Schwinger boson mean field theory has 
some limitations, we hope that the present results will motivate the 
application of more rigorous Monte Carlo or exact-diagonalization 
techniques to our model Hamiltonian.

The paper is organized as follows.
In Sec.~\ref{s2} we give a detailed derivation of our spin-pseudospin 
effective lattice Hamiltonian that we obtain by using a functional integral approach. 
In Sec.~\ref{s3} we discuss the generalization of the Schwinger boson 
mean field theory of Arovas and Auerbach to the case of spin models
with anisotropy.  In Sec.~\ref{s4} we apply these considerations to our 
spin-pseudospin model for bilayer quantum Hall systems, discussing 
the results of our calculations and commenting on their relevance to experiments.
We conclude in Sec.~\ref{s5} by briefly discussing the application of our approach
to other broken symmetry states at integer filling factors in quantum 
Hall systems, including the case of $\nu=2$ in bilayers.

\section{Generalized spin model}
\label{s2}

A single-particle state of an electron in a QH bilayer is described by three
quantum numbers: LLL orbital state quantum number $i$, spin $\sigma$ and 
pseudospin $\tau$. 
In the incompressible state at filling factor $\nu=1$
one can assume at low enough temperatures that there are no charge fluctuations and
that each LLL orbital state $i$ is always occupied by exactly one electron.
The model we derive starting from this assumption describes quantum 
fluctuations
of the remaining spin and pseudospin degrees of freedom.
We will allow electrons in each LLL orbital to be in the most general 
coherent superposition of spin and pseudospin eigenstates.
To define our notation, we write the corresponding creation operator as 
\begin{equation}
\label{1}
\psi_i^{\dagger}=\sum_{k=1}^4 z_{ik} c^{\dagger}_{ik}.
\end{equation}
Here we have introduced 4-component spinor index $k=1,...,4$ to describe 
the mixed spin-pseudospin degree of freedom.
$k=1$ labels an up-spin electron in the top layer, 
$k=2$ a down-spin top-layer electron, $k=3$ a bottom-layer up-spin electron,
and $k=4$ a bottom-layer down-spin electron.
$c^{\dagger}_{ik}$ is the creation operator for an electron in the 
LLL orbital state $i$ and with 4-spinor index $k$. 
$z_{ik}$ are complex amplitudes that satisfy the normalization constraint 
\begin{equation}
\label{2}
\sum_{k=1}^4 |z_{ik}|^2 = 1.
\end{equation}
A single Slater determinant many-body wavefunction can be constructed 
by putting exactly one electron with an arbitrary spinor
(\ref{1}) (for $\nu=1$) in every LLL orbital state $i$
\begin{equation}
\label{3}
|\Psi[z]\rangle = \prod_i \left(\sum_{k=1}^4 z_{ik} c^{\dagger}_{ik}
\right)| 0 \rangle.
\end{equation}
Our approach is based on the observation that this set of states is 
complete when charge fluctuations are neglected. 

We assume that the bilayer is described by a Hamiltonian of the following 
form~\cite{Macdonald99}
\begin{eqnarray}
\label{4}
H&=&\sum_{k_1 k_2 i}c^{\dagger}_{i k_1}h^0_{k_1 k_2} c_{i k_2}\nonumber\\
&+&\frac{1}{2}\sum_{k_1 k_2}\sum_{i_1 i_2 i_3 i_4} c^{\dagger}_{i_1 k_1}
c^{\dagger}_{i_2 k_2} c_{i_4 k_2} c_{i_3 k_1} 
\left[\langle i_1 i_2 |V_+| i_3 i_4 \rangle + \tau^z_{k_1k_1}
\tau^z_{k_2k_2}\langle i_1 i_2 |V_-| i_3 i_4 \rangle\right].
\end{eqnarray}
Here we have introduced $4\times4$ Pauli matrices 
$\tau^a$ corresponding to the usual Pauli operators acting on  
pseudospin degrees of freedom only. 
The analogous spin Pauli matrices will be denoted $\sigma^a$.
$V_{\pm}=(V_S \pm V_D)/2$ and $V_S$ and $V_D$ are the 2D Coulomb interactions 
between electrons in the same and different layers respectively.
The single particle part of the Hamiltonian, $h^0$, consists in general 
of three terms, corresponding to an interlayer bias potential, an 
interlayer tunneling amplitude, and Zeeman 
coupling to the perpendicular magnetic field
\begin{equation}
\label{5}
h^0=-\frac{\Delta_V}{2} \tau^z - \frac{\Delta_t}{2} \tau^x - 
\frac{\Delta_z}{2} \sigma^z.
\end{equation}

Since the occupation of each orbital state is fixed
at one for $\nu=1$ in our approach, the orbital, or charge degree of freedom in (\ref{4}) is 
irrelevant.  We would therefore like to completely eliminate any reference to the 
charge degree of freedom present in the 
microscopic Hamiltonian, obtaining an effective Hamiltonian that refers only to
spin and pseudospin variables.  The most convenient and mathematically 
rigorous way to do this is provided by a functional integral approach.
Assuming at each discrete imaginary time that the single Slater determinant 
states are complete, and following standard lines for the derivation of  
path integral
formulations of quantum statistical mechanics problems,
we obtain the following formal expression for the partition function of a 
$\nu=1$ quantum Hall ferromagnet: 
\begin{equation}
\label{6}
Z=\int D[\bar z, z]e^{-S[\bar z, z]},
\end{equation}  
where the imaginary time action is given by
\begin{eqnarray}
\label{7}
&&S[\overline z, z]=\int_0^{\beta} 
d\tau \left[\sum_i(\langle z_i | \partial_{\tau} |z_i \rangle -
\frac{\Delta_V}{2}\langle z_i|\tau^z|z_i\rangle -
\frac{\Delta_t}{2}\langle z_i|\tau^x|z_i\rangle -
\frac{\Delta_z}{2}\langle z_i|\sigma^z|z_i\rangle)\right.\nonumber\\
&+&\left.\frac{1}{2} \sum_{ij}( H_{ij} \langle z_i |\tau^z|z_i\rangle 
\langle z_j |\tau^z|z_j\rangle- 
F^+_{ij} \langle z_i|z_j\rangle \langle z_j|z_i
\rangle - F^-_{ij} \langle z_i|\tau^z|z_j\rangle 
\langle z_j|\tau^z|z_i\rangle) \right].
\end{eqnarray} 
Here $H_{ij}=\langle ij|V_-|ij\rangle$, and $F^{\pm}_{ij}=\langle ij|V_{\pm}|ji
\rangle$ are the direct and exchange two-particle matrix elements of the 
Coulomb interaction.

Using this path integral representation as an intermediary, we can 
now write the Hamiltonian in terms of spin and pseudospin operators.
The complex fields in this path integral can be thought of as 
bosonic coherent state variables and 
we can introduce bosonic creation operators $a^{\dagger}_{ik}$
corresponding to complex variables $z_{ik}$. 
The normalization condition (\ref{2}) translates into the 
single-occupancy constraint for the $a$ bosons 
\begin{equation}
\label{8}
\sum_{k=1}^4 a^{\dagger}_{ik} a_{ik}=1.
\end{equation}
The action (\ref{7}) is identical to that which arises from a coherent state 
path integral representation of the partition function for a system of bosons
described by the following Hamiltonian
\begin{eqnarray}
\label{9}
&&H=\sum_{ik_1k_2}a^{\dagger}_{ik_1}h^0_{k_1k_2}a_{ik_2}\nonumber\\
&+&\frac{1}{2}\sum_{ij}\sum_{k_1k_2}\left[ H_{ij}a^{\dagger}_{ik_1}
a^{\dagger}_{jk_2}a_{ik_1}a_{jk_2}\tau^z_{k_1k_1}\tau^z_{k_2k_2}-
F^+_{ij}a^{\dagger}_{ik_1}a^{\dagger}_{jk_2}a_{ik_2}a_{jk_1}-
F^-_{ij}a^{\dagger}_{ik_1}
a^{\dagger}_{jk_2}a_{ik_2}a_{jk_1}\tau^z_{k_1k_1}\tau^z_{k_2k_2}\right].
\end{eqnarray}
One can think of the bosons we introduce as Schwinger bosons representing
``spins'' which are generators of the SU(4) group (the usual physical 
spins are SU(2) generators). 
The SU(4) generators are written in terms of the Schwinger boson creation
and annihilation operators as follows
\begin{equation}
\label{10}
S^{k_1k_2}_i=a^{\dagger}_{ik_1}a_{ik_2}.
\end{equation}
The Hamiltonian (9) can now be rewritten in terms of the SU(4) ``spin''
operators $S^{k_1k_2}_i$
\begin{eqnarray}
\label{11}
H&=&\sum_{ik_1k_2}S^{k_1k_2}_i h^0_{k_1k_2}\nonumber\\
&+&\frac{1}{2}\sum_{ij}\sum_{k_1k_2}\left[ H_{ij}S^{k_1k_1}_i
S^{k_2k_2}_j\tau^z_{k_1k_1}\tau^z_{k_2k_2}-
F^+_{ij}S^{k_1k_2}_i S^{k_2k_1}_j-
F^-_{ij}S^{k_1k_2}_i S^{k_2k_1}_j\tau^z_{k_1k_1}\tau^z_{k_2k_2}\right].
\end{eqnarray}     
We have thus mapped the original problem of interacting electrons in a 
quantum Hall bilayer onto a generalized ``spin'' problem. 
Each LLL orbital $i$ (``lattice site'' in the spin model language)
is occupied by an SU(4) ``spin'' and  
``spins'' on different sites are coupled via a long-range anisotropic
interaction (we will rewrite these spin operators below in terms of more 
physically transparent SU(2) spin and pseudospin operators). 

So far we have not specified the LLL orbital basis we are using.
In fact, the accuracy of the model that follows from neglecting charge 
fluctuations does depend on the single-particle representation we employ.
It is obvious that the usually used orbit-center quantum numbers in Landau or
symmetric gauges are not good choices for the complete set of orbital labels
since they are not localized and the energy penalty for double occupation
will vanish in the thermodynamic limit.  Our approach is best used in 
combination with a Wannier-like von Neumann lattice basis set for
a Landau level,~\cite{Ferrari,Efros} in which orbitals are centered on 
lattice sites and the 
unit cell area equals the area per flux quantum, $2 \pi \ell^2$, to accomodate
exactly one electron with a corresponding SU(4) ``spin'' per site at $\nu=1$.  
We choose a square lattice with lattice constant $\sqrt{2\pi \ell^2}$.
It is not immediately obvious that such a basis exists.
A strong magnetic field imposes certain restrictions on the localization 
properties of magnetic orbitals,~\cite{Perelomov,Zak,Thouless} 
and it is well established, for example,
that a set of linearly independent {\it and} exponentially localized 
single-particle orbitals in the LLL does not exist.  
However, it turns out to be possible~\cite{Ferrari,Efros} to construct
a complete orthonormal set of Wannier-like eigenfunctions, which,
although not exponentially localized, have a well defined Gaussian core
and a power law fall off at large distances. 
Following~\cite{Ferrari,Efros} we refer to these orthonormal basis states 
as magnetic Wannier functions. 
The procedure one uses to construct such a basis set is very much like
the one used to construct the usual Wannier functions in a crystal. 
One starts from the set of minimum uncertainty wavepackets 
for electrons in the LLL, centered at the sites of the square lattice
described above.  The difference from the case of a crystal here is that this 
set is {\it overcomplete}, as shown by Perelomov.~\cite{Perelomov}
One then constructs Bloch functions from linear combinations of the 
minimum uncertainty wavepackets and Fourier transforms them to obtain 
the Wannier functions. 
There are subtleties in this procedure and 
we refer the reader to the appendix of our paper and to the original 
papers~\cite{Ferrari,Efros} for further details.  For the following discussion
it is important to note that the procedure used to construct the basis states 
depends on the size of the system, so that the spin and 
pseudospin interactions we discuss below have values that depend on the overall
system size, converging to well defined values in the thermodynamic limit. 

The ``spin model'' (\ref{11}) is not very useful as it is because we do not
have much intuition about SU(4) ``spins''.
However, it turns out to be possible to further rewrite (\ref{11}) in a 
much more useful form.
To proceed one notes that the set of SU(4) generators (\ref{10}) 
is not unique.~\cite{Arovas95}
We define a different set of generators, more appropriate for our
purpose, as follows. 
Define total spin and pseudospin operators on each lattice site
\begin{eqnarray}
\label{12}
{\bf S}_i&=&\frac{1}{2}\sum_{k_1k_2}a^{\dagger}_{ik_1}\vec \sigma_{k_1k_2}
a_{ik_2}\nonumber\\
{\bf T}_i&=&\frac{1}{2}\sum_{k_1k_2}a^{\dagger}_{ik_1}\vec \tau_{k_1k_2}
a_{ik_2}.
\end{eqnarray}
It is then possible to show that the original SU(4) generators given by 
(\ref{10})
can be expressed in terms of a new set of generators $\{S^a,T^b,S^aT^b\}$
as follows
\begin{eqnarray}
\label{13}
S^{11}_i&=&\frac{1}{4}+\frac{1}{2}(S^z_i+T^z_i)+S^z_iT^z_i\nonumber\\
S_i^{22}&=&\frac{1}{4}-\frac{1}{2}(S^z_i-T^z_i)-S^z_iT^z_i\nonumber\\
S^{33}_i&=&\frac{1}{4}+\frac{1}{2}(S^z_i-T^z_i)-S^z_iT^z_i\nonumber\\
S^{44}_i&=&\frac{1}{4}-\frac{1}{2}(S^z_i+T^z_i)+S^z_iT^z_i\nonumber\\
S^{12}_i&=&\frac{1}{2}S^+_i+S^+_iT^z_i\nonumber\\
S^{13}_i&=&\frac{1}{2}T^+_i+S^z_iT^+_i\nonumber\\
S^{14}_i&=&S^+_iT^+_i\nonumber\\
S^{23}_i&=&S^-_iT^+_i\nonumber\\
S^{24}_i&=&\frac{1}{2}T^+_i-S^z_iT^+_i\nonumber\\
S^{34}_i&=&\frac{1}{2}S^+_i-S^+_iT^z_i.
\end{eqnarray}
Now the SU(4) ``spin'' Hamiltonian (\ref{11}) can be rewritten,
up to a constant, as the following coupled spin-pseudospin Hamiltonian 
\begin{eqnarray}
\label{14}
H&=&-\sum_i\left[\Delta_V T^z_i+\Delta_t T^x_i+\Delta_z S^z_i\right]
\nonumber\\
&+&\sum_{ij} \left[ (2H_{ij}-\frac{1}{2}F^S_{ij})T^z_iT^z_j-
\frac{1}{2}F^D_{ij}{\bf T}^{\perp}_i \cdot {\bf T}^{\perp}_j-
\frac{1}{2}F^S_{ij}{\bf S}_i \cdot {\bf S}_j\right.\nonumber\\
&-&\left.2 F^S_{ij}({\bf S}_i \cdot {\bf S}_j) T^z_i T^z_j-
2F^D_{ij}({\bf S}_i \cdot {\bf S}_j)({\bf T}^{\perp}_i \cdot {\bf T}^{\perp}_j)\right].
\end{eqnarray}
Here $F^{S,D}=F^+\pm F^-$.

This Hamiltonian (\ref{14}) is one of the main results of our paper.
Let us emphasize that this Hamiltonian is {\it exact}, given the 
assumed absence of charge fluctuations.
It is written in terms of the usual physical spin-$\frac{1}{2}$
operators and therefore all the great variety of methods available for
spin models can be applied to it. 
Note also the quite unexpected simplicity of (\ref{14}). 
All the effects of spin-pseudospin interaction are contained in 
the two terms with only four spin-$\frac{1}{2}$ operators.

The {\it classical} ground state of (\ref{14}) is one with uniform  
spin-polarization in the direction of the applied Zeeman field 
and uniform pseudospin that, for $\Delta_{t}=0$ and $\Delta_{V}=0$,
is in an arbitrary direction in the $\hat x - \hat y$ plane.  It is 
{\em identical} to the Hartree-Fock ground state of the original microscopic Hamiltonian. 
Due to the anisotropic character of (\ref{14}) there will be corrections 
to the Hartree-Fock picture even at zero temperature.
These corrections can be expected to be quite strong, since we are dealing with
spin-$\frac{1}{2}$ operators, i.e. the system is far away from the 
semiclassical regime.  In fact, as we comment later, corrections to 
Hartree-Fock theory grow in importance as the layer separation is increased.
Although these quantum fluctuations can, and have,~\cite{Yogesh01} been 
addressed
in a fully microscopic model at the generalized-random-phase-approximation 
(GRPA) level, the explicit removal of charge degrees of freedom in the 
approximate Hamiltonian we have derived allows a wider variety of more 
powerful
theoretical techniques to be applied.  In particular, the GRPA corresponds
to a linearized spin-wave theory approximation to the above Hamiltonian.  
Later we will go beyond this level approximation, by applying 
Schwinger boson mean field theory to our model Hamiltonian.
 
\begin{figure}
\begin{center}
\epsfxsize=4in
\epsffile{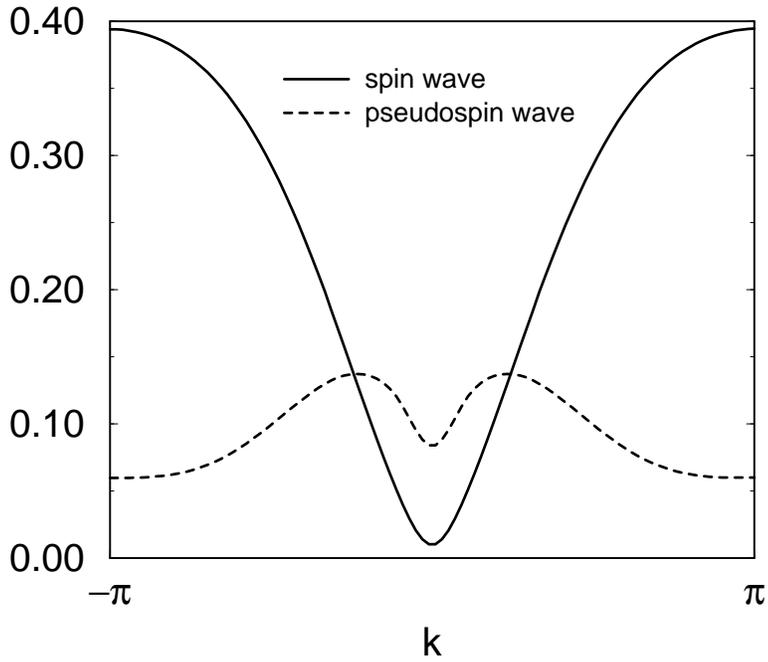}
\caption{Spin and pseudospin wave dispersions for a 20$\times$20 lattice 
and $d/\ell=1.4$ in the (1,1) direction. The dip in the pseudospin wave 
dispersion at the Brillouin zone boundary signals the development of 
antiferromagnetic instability which destroys the long-range XY-ferromagnetic 
order.}
\label{swd}
\end{center}
\end{figure}

We now discuss some simple properties that can be used to simplify 
(\ref{14}) before embarking  on a more detailed study.
We note that $H_{ii}=F^-_{ii}$, {\it i.e.} that on-site direct and exchange Coulomb
matrix elements are identical.  It follows that the on-site contributions
to the Hamiltonian add to a constant term that has no effect on spin and pseudospin dynamics,
a property we use below.  We also note that  
in the absence of the single-particle terms, the Hamiltonian 
has the correct $SU(2)_{spin} \times U(1)_{pseudospin}$ symmetry. 
Correspondingly there are two Goldstone modes associated with the
spontaneous breaking of these symmetries.  Their dispersion relations are 
readily evaluated in the linear spin wave approximation.  For 
$\Delta_{V} =0$ we find that 
\begin{eqnarray}
\label{15}
E_k^{spin}&=&\Delta_z+F^+_0-F^+_k\nonumber\\
E_k^{pseudospin}&=&\sqrt{(\Delta_t+F^D_0+H_k-F^+_k)^2-(H_k-F^-_k)^2}.
\end{eqnarray}
In Fig.~\ref{swd} we show dispersions (\ref{15}) evaluated for a 
20$\times$20 square 
lattice in the $(1,1)$ direction using $\Delta_t=\Delta_z=0.01$ and 
$d/\ell=1.4$.
All energies in this paper will be in units of $e^2/\epsilon \ell$, the 
characteristic energy scale for all fractional quantum Hall systems.
Note that in this example, which corresponds to reasonably typical values,
the pseudospin gap is appreciably 
larger than the spin gap, even though the bare values of tunneling and 
Zeeman coupling are the same.   This difference is due to pseudospin 
fluctuations present in the ground state, even in a linearized spin-wave
theory, and originates from the fact that the Hamiltonian is invariant 
under all spin rotations, but only under pseudospin rotations around the 
$\hat z$ axis, the U(1) symmetry referred to above.  
The dip in the pseudospin wave dispersion at the boundary of the 
Brillouin zone signals the emerging development of antiferromagnetic 
instability,
which eventually destroys the long-range pseudospin ferromagnetic order.
It is this feature of the microscopic physics that is missing in 
long-wavelength effective models. 
 
\begin{figure}
\begin{center}
\epsfxsize=4in
\epsffile{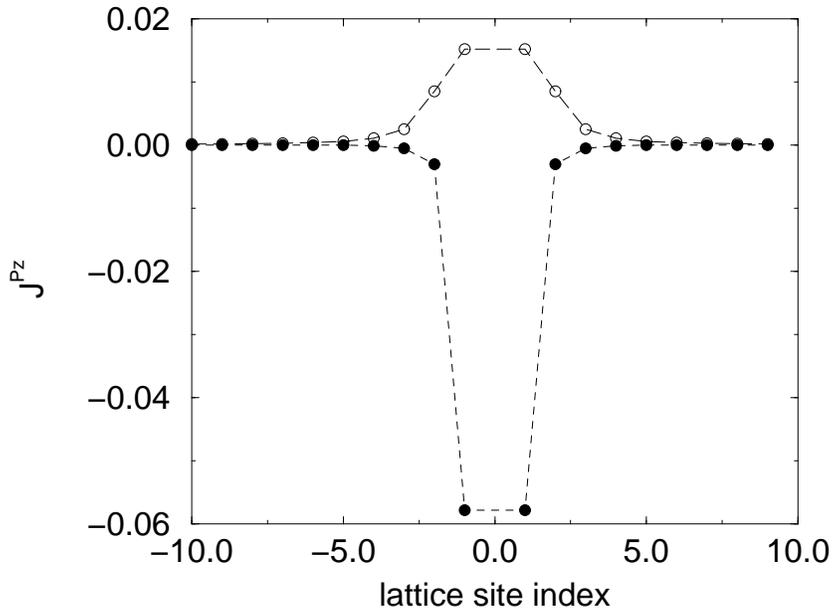}
\caption{Effective pseudospin-pseudospin interaction $J^{Pz}_{ij}$ 
in Eq.(\ref{16}) for 
$d/\ell=0.5$ (filled circles) and $d/\ell=1.4$ (open circles). 
$J^{Pz}_{ij}$ changes its character from short-range ferromagnetic 
at $d/\ell=0.5$ to long-range antiferromagnetic at $d/\ell=1.4$, eventually 
making XY-ferromagnetic pseudospin state unstable.}
\label{coupling1}
\end{center}
\end{figure}

\begin{figure}
\begin{center}
\epsfxsize=4in
\epsffile{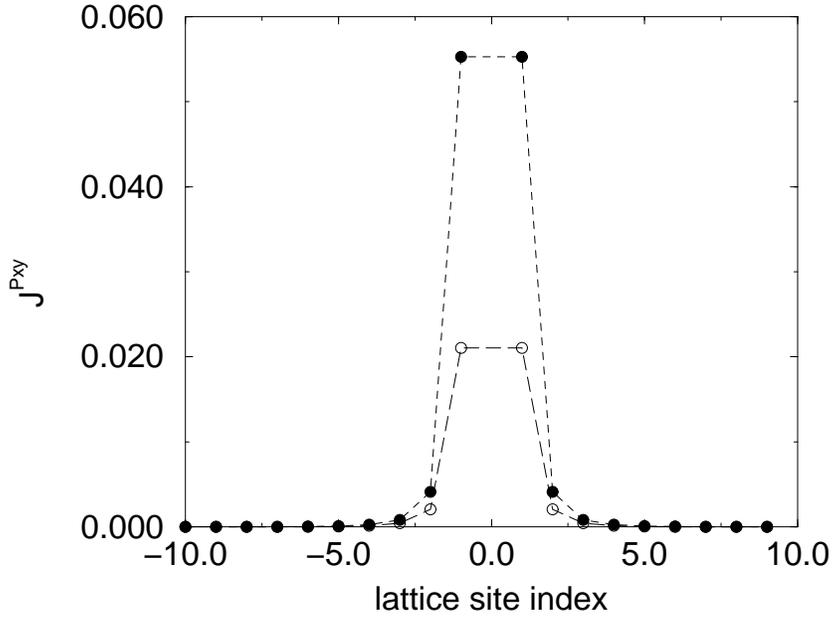}
\caption{Effective pseudospin-pseudospin interaction in 
Eq.(\ref{16}) $J^{P\perp}_{ij}$ for 
$d/\ell=0.5$ (filled circles) and $d/\ell=1.4$ (open circles).
Pseudospin XY-ferromagnetic interactions are weakening as the interlayer 
separation is increased due to the weakening of interlayer exchange 
interactions.}
\label{coupling2}
\end{center}
\end{figure}

Let us now briefly consider the simplified Hamiltonian in which spins
are frozen out (${\bf S}_i \cdot  {\bf S}_j=S^z_i S^z_j=1/4$), valid at $T=0$ 
in the presence of a Zeeman field: 
\begin{equation}
\label{16}
H=-\sum_i\left[\Delta_V T^z_i+\Delta_t T^x_i\right]+
\sum_{ij}\left[(2H_{ij}-F^S_{ij})T^z_iT^z_j-
F^D_{ij}{\bf T}^{\perp}_i \cdot {\bf T}^{\perp}_j\right].
\end{equation}
Consider the evolution of (\ref{16}) as one increases the interlayer 
separation from 0.
At~$d=0$, $2H_{ij}-F^S_{ij}=-F^D_{ij}$ and the system in the absence 
of tunneling and bias is an SU(2) invariant pseudospin ferromagnet.
For $0 < d < d^*$, the $F_{ij}^D$ term dominates and the bilayer ground state 
is an easy-plane pseudospin ferromagnet. 
At a certain critical value of the interlayer separation $d^*$ 
a quantum phase transition occurs to a state with no long-range easy-plane 
ferromagnetic order.
To understand more deeply why this occurs we 
plot in Fig.\ref{coupling1} and \ref{coupling2} the 
pseudospin couplings $J_{ij}^{Pz}=
2H_{ij}-F^S_{ij}$ and $J_{ij}^{P\perp}=F^D_{ij}$ for two different values
of the interlayer separation.  
As can be seen from the figures, increasing the interlayer 
separation primarily affects $J^{Pz}_{ij}$ coupling, 
changing its character from short-range ferromagnetic at $d/\ell=0.5$ to
long-range ($\propto 1/r^3$) antiferromagnetic at $d/\ell=1.4$ (we use 
the word {\it antiferromagnetic} here just to emphasize the sign change 
of the interaction, it is not immediately clear that the ground state 
is actually antiferromagnetic, although it very likely is).
The change in $J_{ij}^{Pz}$ is accompanied by a weakening of the in-plane ferromagnetic interaction $J^{P\perp}_{ij}$.
At present, experimental samples have a $d/\ell$ value that is not 
very far below the critical value.  We therefore expect that the 
effective interactions illustrated for the $d/\ell=1.4$ case are representative
of the current typical experimental situation. 
This change in character reflects a change in the relative importance
of exchange interactions within the layers ($F^S_{ij}$), which are not $d$ 
dependent and the electrostatic interactions between the dipoles ($H_{ij}$) 
that are created by $\hat z$ direction pseudospin polarization and whose
moments are proportional to $d$.  For $d/\ell \sim 1.5$, electrostatics 
dominates, $J_{ij}^{Pz}$ is antiferromagnetic, and competition develops
between two qualitatively different potential ground states.
{\em Our analysis makes it clear 
that it is this developing competition that is responsible 
for the phase transition that occurs in bilayer quantum Hall systems 
for $d/\ell \sim 1.5$.} 
In our model which neglects charge fluctuations, the ground state at large
$d$ is likely an antiferromagnet with Ising anisotropy, as it appears from the 
analysis of the classical energetics of Eq.(\ref{16}).  
In experiment, the loss of the quantum Hall effect at this transition 
indicates that 
it is accompanied by the loss of a charge gap, making charge fluctuations 
at least somewhat important.
However, even if we neglect the charge fluctuations, i.e. still assume 
that the system can 
be described by the pseudospin model (\ref{16}) in the compressible region, 
the precise nature of the pseudospin state above the transition is not 
immediately obvious, although as we have pointed out above, it is likely 
to be antiferromagnetic. 
The $T_i^zT_j^z$ term generally favors an easy axis antiferromagnetic 
ordering of pseudospins. 
However, the long-range character of the $T_i^zT_j^z$ coupling
(since it orginates from Coulomb interactions between dipoles associated with 
$\hat z$-direction pseudospin polarization,
one can expect it to fall off as $1/r^3$) could introduce frustration 
and a nontrivial spin liquid state.
A recent Quantum Monte Carlo study~\cite{Troyer01} of a quantum XXZ model 
with nearest- and next-nearest-neighbor interactions, which is similar 
to (\ref{16}), has shown that both Neel and striped phases are possible,
depending on the relative strength of the interactions.  
Our model could in principle produce more exotic phases due to the truly 
long-range character of spin-spin interactions in our case.
The phase transition between the XY-ferromagnetic and Neel or striped
phases in~\cite{Troyer01} was shown to be first order, which 
gives additional support to the validity of our model, since according to 
recent exact diagonalization studies,~\cite{Schliemann01} 
the compressible-incompressible transition in $\nu=1$ bilayers is very likely 
of first order as well. 
We believe that the pseudospin Hamiltonian 
(\ref{16}) captures
important parts of the physics of this still poorly understood 
quantum phase transition and deserves further study.
  
Although we concentrate in this paper on the case of bilayer quantum Hall systems at 
$\nu=1$ in the lowest Landau level, the same formalism can be used to 
derive spin-pseudospin models for other cases of interest.
For example, pseudospin stripe states~\cite{dlstriperefs}
are expected at $\nu=1$ in the case of Landau levels with orbital index 
$N > 2$, and Ising ferromagnets are expected~\cite{Jungwirth00} in bilayers 
when the orbital indices in the two layers differ. 
In our formalism these different
phases will appear naturally because of changes in the effective pseudospin-pseudospin
interactions.  The effective interactions we calculate depend slightly on the 
finite size of the system in which the magnetic Wannier states are orthogonalized,
but approach a definite value in the thermodynamic limit.  Effective interactions 
for either the thermodynamic limit, or for any specified finite size system, 
are available from the authors on request for any of the above cases.

To investigate finite-temperature properties of our spin-pseudospin 
model for lowest Landau level bilayers,
one needs to go beyond linear spin wave theory (\ref{15}) which 
is not adequate in two space dimensions (2D) at any finite temperature 
due to the absence of long-range order.
A simple but powerful method, the Schwinger boson mean field theory (SBMFT), 
was proposed by Arovas and Auerbach.~\cite{Arovas88}
It does not break spin rotational invariance at finite temperatures, yet
correctly reproduces linearized spin wave theory in the semiclassical limit.
For the SU(2) Heisenberg model SBMFT qualitatively reproduces the low 
temperature continuum field theory results.~\cite{Haldane83,CHN89}
For 1D integer antiferromagnetic spin chains, SBMFT gives the correct value
of the Haldane gap in the ground state.  However, it fails in the case of the 
half-integer chains, where Berry phase effects missed by SBMFT 
result in a gapless excitation spectrum at zero temperature. 

In the next section we briefly recount some features of the original theory,
developed for the SU(2) invariant Heisenberg model, that will be important 
for our discussion.  We then show how to achieve a proper generalization
of SBMFT for spin models with anisotropy, a procedure that requires some care.
We then apply our generalized SBMFT to the model Hamiltonian derived 
in the present section.     

\section{Schwinger boson mean field theory for spin models with anisotropy}
\label{s3}
We first briefly review the original SBMFT proposed by 
Arovas and Auerbach~\cite{Arovas88} for the SU(2) invariant spin model.
Our motivation here is to lay the ground work for the generalization to 
anisotropic interaction models that is required for the application to our 
model Hamiltonian.  
Consider the 2D Heisenberg model on a 
square lattice with only nearest-neighbor interactions
\begin{equation}
\label{17}
H=-\frac{1}{2}J\sum_{<ij>}{\bf S}_i \cdot {\bf S}_j.
\end{equation}
The sum is over neighboring {\it sites} with the $1/2$ factor correcting for 
double counting.  For $J>0$ the ground state of this model is ferromagnetic.
In 2D long-range order is destroyed at any finite temperature,
strong short-range ferromagnetic (or antiferromagnetic for $J<0$) correlations
are present for temperatures up to $\sim |J|$, however.
A way to build a simple theory of this {\em spin liquid} state
uses the Schwinger boson representation of spin operators
\begin{eqnarray}
\label{18}
&&S_i^z=\frac{1}{2}(a^{\dagger}_{i1} a_{i1}-a^{\dagger}_{i2} a_{i2})\nonumber\\
&&S_i^+=a^{\dagger}_{i1}a_{i2}\nonumber\\
&&S_i^-=a^{\dagger}_{i2}a_{i1}\nonumber\\
&&a^{\dagger}_{i1}a_{i1}+a^{\dagger}_{i2}a_{i2}=2S.
\end{eqnarray}
Correlations in the spin liquid are then described by 
{\it bond operators}, which are 
bilinear forms in Schwinger boson operators from nearest-neighbor sites.
Nonzero expectation values for these bond operators represent 
{\it short-range order}.
There are of course many possible bond operators one can define, however 
{\it only
the SU(2) invariant ones are relevant}, since rotational invariance is preserved 
in the spin liquid state. 
There are only two SU(2) invariant bond operators:
\begin{eqnarray}
\label{19}
F^{\dagger}_{ij}&=&a^{\dagger}_{i1}a_{j1}+a^{\dagger}_{i2}a_{j2}\nonumber\\
A^{\dagger}_{ij}&=&a^{\dagger}_{i1}a^{\dagger}_{j2}-
a^{\dagger}_{i2}a^{\dagger}_{j1}.
\end{eqnarray}   
The operators $F^{\dagger}$ and $A^{\dagger}$ are not independent, but
are connected by the following operator identity
\begin{equation}
\label{20}
:F^{\dagger}_{ij}F_{ij}:+A^{\dagger}_{ij}A_{ij}=4S^2,
\end{equation}
where colons represent normal ordering.
One can write the spin Hamiltonian (\ref{17}) in terms of these bond 
operators as follows
\begin{equation}
\label{21} 
H=-\frac{J}{8}\sum_{<ij>}\left(:F^{\dagger}_{ij}F_{ij}:-A^{\dagger}_{ij}
A_{ij}\right)
\end{equation}
From (\ref{21}) the physical meaning of the two bond operators is clear.
Nonzero expectation value of $F^{\dagger}$ represents short-range 
ferromagnetic order, of $A^{\dagger}$---short-range antiferromagnetic order.
Using (\ref{20}) one can eliminate one of the bond operator products from 
(\ref{21}), depending on the kind of correlations one expects to have 
(determined by the sign of J in the simple Heisenberg model).
Assuming $J > 0$ one has
\begin{equation}
\label{22}
H=-\frac{J}{4}\sum_{\langle ij\rangle}
\left(:F^{\dagger}_{ij}F_{ij}:-2S^2\right).
\end{equation}

In the original approach of Arovas and Auerbach 
this representation is generalized to $N$ Schwinger boson flavors,
making it possible to set up a systematic expansion of the functional
integral representation of the partition function     
in powers of $1/N$, starting from the saddle point approximation. 
We are, however, only interested in the saddle point, or mean field, 
solution itself and therefore will use a more simple-minded approach.
We assume that the bond operator has a nonzero expectation value 
\begin{equation}
\label{23}
Q = \langle F^{\dagger}_{ij}\rangle = \langle F_{ij}\rangle,
\end{equation}
and perform a Hartree-Fock decoupling of (\ref{22}).
The mean field Hamiltonian, neglecting constants, is
\begin{eqnarray}
\label{24}
H&=&-\frac{JQ}{4}\sum_{<ij>}\left(F^{\dagger}_{ij}+F_{ij}\right)\nonumber\\
&=&-\frac{JQ}{4}\sum_{<ij>}\left(a^{\dagger}_{i1}a_{j1}+
a^{\dagger}_{i2}a_{j2}+h.c.\right).
\end{eqnarray} 
The constraint on Schwinger boson occupation numbers in (\ref{18}) is imposed 
on average by introducing a chemical potential term in the mean field 
Hamiltonian to yield  
\begin{equation}
\label{25}
H=\lambda\sum_i\left(a^{\dagger}_{i1}a_{i1}+a^{\dagger}_{i2}a_{i2}\right)-
\frac{JQ}{4}\sum_{<ij>}\left(a^{\dagger}_{i1}a_{j1}+
a^{\dagger}_{i2}a_{j2}+h.c.\right),
\end{equation}
or in Fourier space
\begin{equation}
\label{26}
H=\sum_{\bf k} \epsilon_{\bf k}(a^{\dagger}_{{\bf k} 1}a_{{\bf k} 1}+
a^{\dagger}_{{\bf k} 2}a_{{\bf k} 2}),
\end{equation}
where
\begin{equation}
\label{27}
\epsilon_{\bf k}=\lambda-2JQ\gamma_{\bf k},
\end{equation}
and $\gamma_{\bf k}=\frac{1}{2}(\cos(k_xa)+\cos(k_ya))$ for a square 
lattice with lattice constant $a$.
It is convenient to redefine $\lambda$ to make the notation more physically 
meaningful, and also to facilitate the solution of SBMFT equations by 
constraining the form of $\epsilon_{\bf k}$
\begin{equation}
\label{28}
\lambda \rightarrow \lambda+2JQ.
\end{equation}
Then $\epsilon_{\bf k}$ becomes
\begin{equation}
\label{29}
\epsilon_{\bf k}=\lambda+2JQ(1-\gamma_{\bf k}).
\end{equation}
One completes the model by writing down the selfconsistency equations
\begin{eqnarray}
\label{30}
&&\frac{2}{N}\sum_{\bf k}n_B(\epsilon_{\bf k})=2S\nonumber\\
&&Q=2S-\frac{2}{N}\sum_{\bf k}(1-\gamma_{\bf k})n_B(\epsilon_{\bf k}),
\end{eqnarray}
where $N$ is the total number of lattice sites.

The first of equations (\ref{30}) is the constraint on the {\it average}
number of Schwinger bosons per site, while the second is the self-consistent
expression for the bond operator expectation value. 
The physical meaning of $\epsilon_{\bf k}$ is now clear.
The second term in (\ref{29}) is just the gapless magnon dispersion with
a self-consistently renormalized stiffness constant. 
The gap $\lambda$ reflects the finite correlation length in the spin 
liquid at finite temperatures.
As the temperature goes to zero and the system size $N$ goes to infinity, 
$\lambda \rightarrow 0$, $Q \rightarrow 2S$ and 
the gapless quadratic spin wave dispersion is recovered.
Note however, that the mean field quasiparticles in SBMFT are {\it not} 
magnons.
Due to the fact that the constraint on the total number of Schwinger 
bosons on each site is enforced in (\ref{30}) only on average, unphysical
excitations which change the local number of Schwinger bosons are
allowed in the mean field theory.
 
It is straightforward to show that at finite temperatures SBMFT reproduces 
the {\it renormalized classical}~\cite{CHN89} correlation 
length for 2D quantum ferromagnets and antiferromagnets.  
For 1D integer antiferromagnetic chains at $T=0$, the Haldane gap is also
correcty described.
 
Generally, SBMFT has proven to be a very successful and useful tool. 
One can think of it as an analog of the Weiss molecular field theory for 
quantum spin liquids.   
It is accurate enough to describe subtle quantum effects, yet simple enough to
treat complicated spin models, such as (\ref{14}). 
We refer the reader to the original paper of Arovas and Auerbach and 
Auerbach's textbook~\cite{Auerbach} for 
further details on the application of SBMFT to the SU(2) Heisenberg model.    
For a discussion of the application of the SBMFT approach to
frustrated antiferromagnets see.~\cite{Chandra} 

We now generalize the theory outlined above to anisotropic models.
It is clear how to do this if one recalls the crucial principle of
the SBMFT: one needs to write the Hamiltonian in terms of products of bond 
operators invariant under the transformations of the symmetry group
of the Hamiltonian.  To be concrete, let us consider the following model
\begin{equation}
\label{31}
H=\frac{1}{2}\sum_{<ij>}\left(J^zS^z_iS^z_j-J^{\perp}{\bf S}^{\perp}_i
\cdot {\bf S}^{\perp}_j
\right).
\end{equation}
This Hamiltonian has the same $Z_2\times U(1)$ symmetry as the pseudospin 
part of our spin-pseudospin model (\ref{14}) and shares the frustration that 
can occur between antiferromagnetic $S^z-S^z$ interactions and ferromagnetic
$S^{\perp}-S^{\perp}$ interactions.
To illustrate our strategy we assume a square lattice with only nearest neighbor spins interacting.
We will also discuss only the case $J^{\perp} > J^z$ when the 
classical ground state is ferromagnetically ordered in the 
$\hat x - \hat y$ plane, 
which is the case relevant case for bilayer quantum Hall ferromagnets.
Other cases can be analyzed in a similar way.  

There are four bond operators that are invariant under symmetry operations of 
the Hamiltonian:
\begin{eqnarray}
\label{32}
&&F^{\dagger}_{ij}=a^{\dagger}_{i1}a_{j1}+a^{\dagger}_{i2}a_{j2}\nonumber\\
&&A^{\dagger}_{ij}=a^{\dagger}_{i1}a^{\dagger}_{j2}-
a^{\dagger}_{i2}a^{\dagger}_{j1}\nonumber\\
&&X^{\dagger}_{ij}=a^{\dagger}_{i1}a^{\dagger}_{j2}+
a^{\dagger}_{i2}a^{\dagger}_{j1}\nonumber\\
&&Z^{\dagger}_{ij}=a^{\dagger}_{i1}a_{j1}-a^{\dagger}_{i2}a_{j2}.
\end{eqnarray}
The first two are the usual SU(2) ferromagnetic and antiferromagnetic 
bond operators.
The two new bond operators, $X^{\dagger}_{ij}$ and $Z^{\dagger}_{ij}$ 
represent XY and easy axis ferromagnetic correlations.
They are connected by the familiar identity 
\begin{equation}
\label{32.1}
:Z^{\dagger}_{ij}Z_{ij}:+X^{\dagger}_{ij}X_{ij}=4S^2.
\end{equation}

As in the SU(2) Heisenberg model case, one can write the Hamiltonian
in terms of the products of invariant bond operators. 
There are again several ways to do this, and we
need to choose the one that is appropriate to describe the correlations in the 
ground state.  In the case where the ground state is an XY-ferromagnet, the 
desired mean field theory follows from the following form for the 
Hamiltonian 
\begin{equation}
\label{33}
H=\frac{1}{8}\sum_{<ij>}\left[J^z\left(:F^{\dagger}_{ij}F_{ij}:-
X^{\dagger}_{ij}X_{ij}\right)-J^{\perp}
\left(X^{\dagger}_{ij}X_{ij}+:F^{\dagger}_{ij}F_{ij}:-4S^2\right)\right].
\end{equation}
We will see later that the resulting Schwinger boson mean field theory
correctly reproduces quasiclassical dynamics in the $S\rightarrow \infty,
T\rightarrow 0$ limit.
 
Regrouping terms and neglecting a constant contribution, the Hamiltonian 
can be rewritten as  
\begin{equation}
\label{34}
H=-\frac{1}{8}\sum_{<ij>}\left[(J^{\perp}-J^z):F^{\dagger}_{ij}F_{ij}:+
(J^{\perp}+J^z)X^{\dagger}_{ij}X_{ij}\right].
\end{equation}      
To proceed it is convenient to first rotate coordinates by $\pi/2$ around 
the y-axis.
The $X^{\dagger}$ bond operator then changes to
\begin{equation}
\label{35}
X^{\dagger}_{ij}=a^{\dagger}_{i1}a^{\dagger}_{j1}-a^{\dagger}_{i2}
a^{\dagger}_{j2}.
\end{equation}
$F^{\dagger}_{ij}$ does not change of course, because it is an SU(2) 
invariant.
As before, we introduce expectation values of the bond operators
\begin{eqnarray}
\label{36}
&&Q=\langle F^{\dagger}_{ij}\rangle=\langle F_{ij} \rangle\nonumber\\
&&P=\langle X^{\dagger}_{ij}\rangle=\langle X_{ij} \rangle, 
\end{eqnarray}
and perform a Hartree-Fock decoupling.  The resulting mean field Hamiltonian is
\begin{eqnarray}
\label{37}
H&=&\sum_{\bf k}\left[\lambda-(J^{\perp}-J^z)Q\gamma_{\bf k}\right]
\left(a^{\dagger}_{{\bf k}1}a_{{\bf k}1}
+a^{\dagger}_{{\bf k}2}a_{{\bf k}2}\right)\nonumber \\
&-&\frac{1}{2}\left(J^{\perp}+J^z\right)P\sum_{\bf k}\gamma_{\bf k}
\left(a^{\dagger}_{{\bf k}1}a^{\dagger}_{{\bf -k}1}+
a_{{\bf k}1}a_{{\bf -k}1}-
a^{\dagger}_{{\bf k}2}a^{\dagger}_{{\bf -k}2}-a_{{\bf k}2}a_{{\bf -k}2}\right).
\end{eqnarray}

We introduce the following notation to make the subsequent equations 
readable
\begin{eqnarray}
\label{37.1}
A_{{\bf k}1}&=&\lambda+(J^{\perp}+J^z)P+(J^{\perp}-J^z)
Q(1-\gamma_{\bf k})\nonumber \\
A_{{\bf k}2}&=&\lambda+h+(J^{\perp}+J^z)P+(J^{\perp}-J^z)
Q(1-\gamma_{\bf k})\nonumber \\
B_{{\bf k}1,2}&=&\pm(J^{\perp}+J^z)P\gamma_{\bf k}\nonumber \\
\epsilon_{{\bf k}1,2}&=&\sqrt{A_{{\bf k}1,2}^2-B_{{\bf k}1,2}^2}.
\end{eqnarray}
Here we have added magnetic field $h$ in the $\hat x$-direction and
redefined $\lambda$ as before.

Eq.(\ref{37}) in this notation becomes
\begin{equation}
\label{37.2}
H=\sum_{{\bf k},m=1,2}\left[A_{{\bf k}m}a^{\dagger}_{{\bf k}m}a_{{\bf k}m}-
\frac{1}{2}B_{{\bf k}m}\left(a^{\dagger}_{{\bf k}m}a^{\dagger}_{{\bf -k}m}+
a_{{\bf k}m}a_{{\bf -k}m}\right)\right].
\end{equation}
Hamiltonian (\ref{37.2}) is diagonalized by a standard Bogoliubov 
transformation
\begin{eqnarray}
\label{37.3}
&&a_{{\bf k}m}=\cosh\theta_{{\bf k}m}\alpha_{{\bf k}m}+\sinh\theta_{{\bf k}m}
\alpha^{\dagger}_{-{\bf k}m}\nonumber\\
&&\cosh2\theta_{{\bf k}m}=\frac{A_{{\bf k}m}}{\epsilon_{{\bf k}m}} \nonumber\\
&&\sinh2\theta_{{\bf k}m}=\frac{B_{{\bf k}m}}{\epsilon_{{\bf k}m}}, 
\end{eqnarray}
and
\begin{equation}
\label{38}
H=\sum_{{\bf k}m}\epsilon_{{\bf k}m}\alpha^{\dagger}_{{\bf k}m}
\alpha_{{\bf k}m}.
\end{equation}
The parameters in the mean field Hamiltonian are fixed by solving the
self-consistency equations 
\begin{eqnarray}
\label{40}
&&\frac{1}{N}\sum_{\bf k}\left\{\frac{A_{{\bf k}1}}{\epsilon_{{\bf k}1}}
\left[n_B(\epsilon_{{\bf k}1})+\frac{1}{2}\right]+
\frac{A_{{\bf k}2}}{\epsilon_{{\bf k}2}}
\left[n_B(\epsilon_{{\bf k}2})+\frac{1}{2}\right]-1\right\}=2S\nonumber\\
&&Q=2S-
\frac{1}{N}\sum_{\bf k}(1-\gamma_{\bf k})
\left\{\frac{A_{{\bf k}1}}{\epsilon_{{\bf k}1}}
\left[n_B(\epsilon_{{\bf k}1})+\frac{1}{2}\right]+
\frac{A_{{\bf k}2}}{\epsilon_{{\bf k}2}}
\left[n_B(\epsilon_{{\bf k}2})+\frac{1}{2}\right]-1\right\}
\nonumber\\
&&P=\frac{1}{N}\sum_{\bf k}\left\{\frac{B_{{\bf k}1}}
{\epsilon_{{\bf k}1}}\left[n_B(\epsilon_{{\bf k}1})+\frac{1}{2}
\right]-
\frac{B_{{\bf k}2}}
{\epsilon_{{\bf k}2}}\left[n_B(\epsilon_{{\bf k}2})+\frac{1}{2}\right]
\right\}.
\end{eqnarray}
 
As can be seen from (\ref{40}), there are corrections to 
linearized spin-wave
theory even in the $T=0$ limit in the anisotropic spin model case.  
This difference is not unexpected since the XY-ferromagnetic order parameter 
does not
commute with the Hamiltonian and has quantum fluctuations in the ground state.  
If both quasiclassical ($S\rightarrow \infty$) and $T\rightarrow 0$ limits are taken, it is easy to see from Eq.(\ref{37.1}) and (\ref{40}) 
that $\lambda\rightarrow 0$,
$Q$ and $P\rightarrow 2S$, and one recovers the magnon dispersion of 
linear spin wave theory.

At low temperatures a system described by (\ref{31}) must have 
{\it quasi long range} order ({\it i.e.} power law correlations) and undergo 
a Kosterlitz-Thouless phase transition at a finite temperature. 
Unfortunately, our theory fails to reproduce these results. 
Instead our SBMFT always implies a nonzero gap in the Schwinger boson spectrum,
and exponential correlation functions, at all finite temperatures. 
This deficiency can be verified by a simple inspection of the
first of equations (\ref{40}),
which represents the constraint on the number of Schwinger bosons, enforced 
on average. 
Assuming $\lambda=0$, the Schwinger bosons dispersion (\ref{37.1}) is linear 
for small $k$.  Taking into account only the leading contribution to the integral over 
wavevector, we can take $n_B(\epsilon_{\bf k}) \sim T/\epsilon_{\bf k}$.
Then it is clear that the integral over ${\bf k}$ in (\ref{40}) diverges 
logarithmically at the origin in 2D.
Therefore at any finite temperature the gap in the Schwinger boson spectrum 
must be nonzero, which translates to exponentially decaying spin correlations  
as the gap introduces a spatial scale into the problem. 
The reason for this failure is the 
presence of unphysical fluctuations in the local number of Schwinger bosons
allowed by the soft (enforced only on average) constraint in (\ref{40}). 
The existense of quasi long range order depends critically on the
fact that the physical low energy excitations in (\ref{31}) (spin waves) are  
mostly in the $\hat x - \hat y$ plane. 
Unphysical fluctuations of the Schwinger boson occupation numbers 
in the mean field theory violate this constraint and therefore destroy the
quasi long range order. 
In the SU(2) case, correlations are not so delicate and therefore
the SBMFT works out qualitatively correctly. 

This deficiency is not a very serious one, however, since in most cases of 
interest a small pseudospin magnetic field is present due to tunneling.
We are concerned mainly with calculating quantities like the 
pseudospin magnetization at finite tunneling and the transverse susceptibility,
which should not be extremely sensitive to the precise character of
long-range correlations.  We have checked the mean field values of the gap in 
the XY-model case are indeed much smaller at all temperatures than the ones in the isotropic 
Heisenberg model.  As is true for any mean field theory, we do not expect our theory to be 
quantitatively correct, but to capture the general trends in the temperature 
and magnetic field dependence of experimentally observable quantities.  
Finally, it seems likely that this problem can be corrected by taking 
into account gaussian fluctuations around the saddle point solution 
(\ref{40}).  Gaussian fluctuations of the field $\lambda$ will reintroduce the local 
constraint and therefore eliminate the unphysical excitations. 
This issue will be examined in the future work.

\begin{figure}
\begin{center}
\epsfxsize=4in
\epsffile{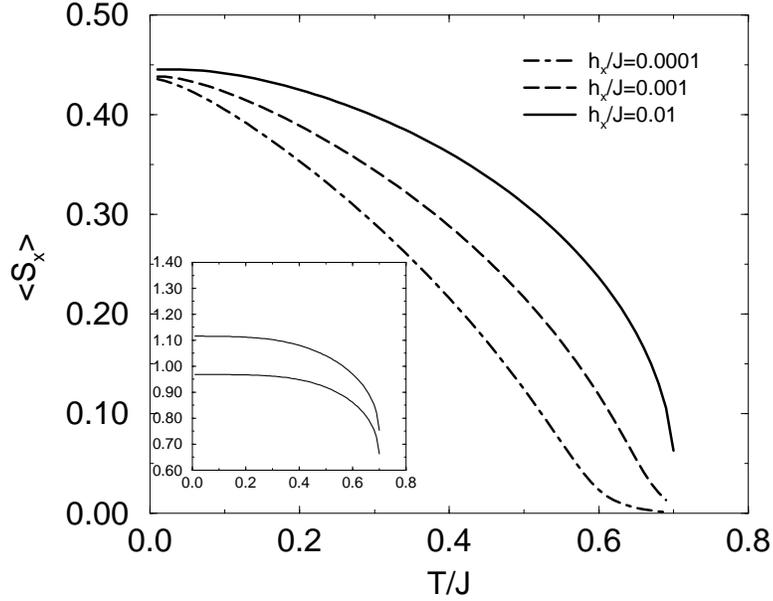}
\caption{In-plane magnetization for the pure XY-model case ($J^z=0$)
of Eq.(\ref{31}).
SBMFT short range order parameters $Q$ and $P$ are shown in the inset.
Magnetization is reduced from its maximum value even at zero temperature due 
to the quantum fluctuations induced by anisotropy.} 
\label{xymag}
\end{center}
\end{figure}

\begin{figure}
\begin{center}
\epsfxsize=4in
\epsffile{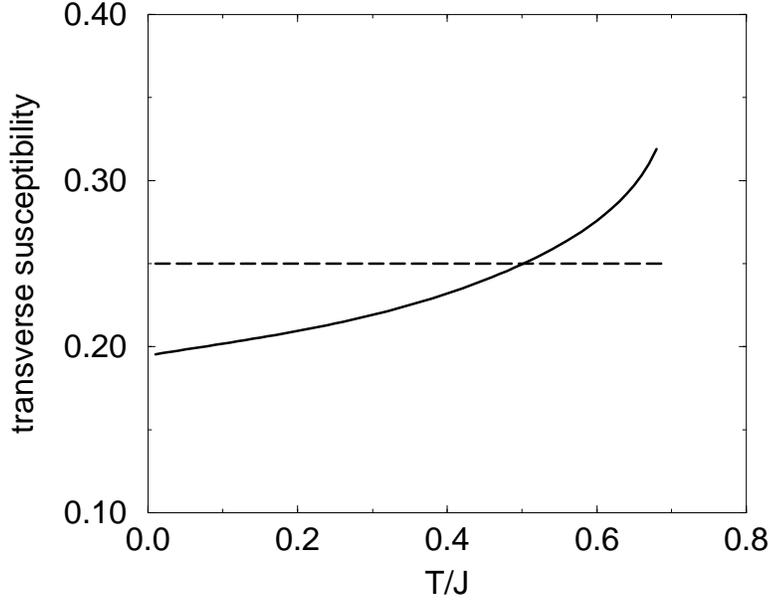}
\caption{Temperature dependence of the transverse susceptibility for the 
XY-model (solid line). 
The classical zero temperature susceptibility is shown by a dashed line. 
Susceptibility increases with temperature due to 
softening of the in-plane spin order by thermal fluctuations.} 
\label{xysusc}
\end{center}
\end{figure}

We have solved the mean field equations (\ref{40}) for 
finite lattices with up to 100$\times$100 sites and periodic boundary 
conditions which make our grid of wavevectors discrete.
Results for the pure XY-model case ($J^z=0$) are summarized in 
Figs.~\ref{xymag} and \ref{xysusc}. 
Fig.~\ref{xymag} shows the in-plane magnetization for different values of the 
magnetic field in the $\hat x$-direction. 
Typical temperature dependences of the short-range order parameters 
$Q$ and $P$ are shown in the inset. 
At a temperature of the order of $J^{\perp}$, there is an unphysical 
transition to the state with no short range correlations.
This is a well known problem of SBMFT which occurs in the SU(2) case as well.

The most interesting quantity for us is the transverse susceptibility
\begin{equation}
\label{chiz0}
\chi^{zz}=\frac{d\langle S^z\rangle}{d h^z}|_{h^z=0}.
\end{equation}
as it is the most relevant quantity for experiments in the case of 
quantum Hall bilayers.
It can be written in terms of imaginary time spin-spin correlation function as
\begin{equation}
\label{chiz}
\chi^{zz}=\frac{1}{N}\sum_{ij}\int_0^{1/T}d{\tau}\langle S^z_i(\tau)
S^z_j(0)\rangle.
\end{equation}
The SBMFT result for Eq. (\ref{chiz}) can be evaluated using the imaginary time
path integral technique. The result is
\begin{eqnarray}
\label{chiz1}
\chi^{zz}&=&\frac{1}{4N}\sum_{\bf k}\left[\left(\frac{A_{{\bf k}1}}
{\epsilon_{{\bf k}1}}\frac{A_{{\bf k}2}}{\epsilon_{{\bf k}2}}+
\frac{B_{{\bf k}1}}
{\epsilon_{{\bf k}1}}\frac{B_{{\bf k}2}}{\epsilon_{{\bf k}2}}+1\right)
\frac{n_B(\epsilon_{{\bf k}2})-n_B(\epsilon_{{\bf k}1})}
{\epsilon_{{\bf k}1}-\epsilon_{{\bf k}2}}\right.\nonumber\\
&+&\left.\left(\frac{A_{{\bf k}1}}
{\epsilon_{{\bf k}1}}\frac{A_{{\bf k}2}}{\epsilon_{{\bf k}2}}+
\frac{B_{{\bf k}1}}
{\epsilon_{{\bf k}1}}\frac{B_{{\bf k}2}}{\epsilon_{{\bf k}2}}-1\right)
\frac{n_B(\epsilon_{{\bf k}1})+n_B(\epsilon_{{\bf k}2})+1}
{\epsilon_{{\bf k}1}+\epsilon_{{\bf k}2}}\right].
\end{eqnarray}
The temperature dependence of the transverse susceptibility for
the case of XY-model with no in-plane field is shown in Fig.\ref{xysusc}. 
The monotonic increase of the susceptibility with temperature reflects 
the softening of the in-plane order due to thermal fluctuations.
At high temperatures $\chi^{zz}$ must follow the usual $1/T$ paramagnetic 
susceptibility temperature dependence.    
Therefore $\chi^{zz}$ must have maximum at an intermediate temperatures
beyond the range over which SBMFT is reliable.
Recent quantum Monte Carlo simulation results~\cite{Sandvik99} support this 
picture. 
Our result for the zero temperature susceptibility $\chi^{zz}=0.1955$  
is reasonably close to the value quantum Monte Carlo result 0.2096 
from.~\cite{Sandvik99}
It is interesting to note that in this case SBMFT apparently gives the correct 
result for the susceptibility, unlike in the SU(2) Heisenberg model case,
where it is overestimated by a factor of $3/2$. 
The temperature dependence we obtain is, however, stronger than 
in~\cite{Sandvik99} 
presumably due to the fact that SBMFT allows unphysical low energy 
out-of-plane spin fluctuations.   

In closing this section, we remark that Timm and Jensen~\cite{Timm00} have
recently suggested a different strategy for generalizing SBMFT to 
anisotropic spin models.  These authors employed a more formal $1/N$ expansion approach. 
We believe that our approach is better for practical calculations since, 
unlike the method of Timm and Jensen, it reproduces the correct semiclassical 
dynamics in the limit $S\rightarrow\infty$ at the mean field level.
         
\section{Schwinger boson mean field theory for the $\nu=1$ bilayer effective 
spin-pseudospin model.}
\label{s4}
In this section we apply the formalism developed above to the effective
spin-pseudospin model of the $\nu=1$ quantum Hall bilayer given by (\ref{14}).
The Hamiltonian (\ref{14}) has an SU(2) invariant spin system coupled 
to a $Z_2\times U(1)$
invariant pseudospin system. 
Therefore our strategy will be to decouple the spin and pseudospin parts 
of (\ref{14}) in a mean field approximation and use
SBMFT to study the decoupled spin and pseudospin Hamiltonians.
The coupling is restored by solving the resulting equations self-consistently.
We expect this to be a good approximation due to the strong short-range 
correlations in our model and the long-range character of the couplings in 
(\ref{14}).

The decoupled spin and pseudospin Hamiltonians are given by
\begin{equation}
\label{41}
H^S=-\Delta_z\sum_iS^z_i
-\sum_{ij}\left[\frac{1}{2}F^S_{ij}+
2F^S_{ij}\langle T^z_i T^z_j\rangle +2F^D_{ij}\langle {\bf 
T}^{\perp}_i \cdot {\bf T}^{\perp}_j\rangle\right]{\bf S}_i \cdot {\bf S}_j,
\end{equation}
and
\begin{eqnarray}
\label{42}
H^P&=&-\sum_i\left[\Delta_V T^z_i+\Delta_t T^x_i\right]\nonumber\\
&+&\sum_{ij}\left[\left(2H_{ij}-\frac{1}{2}F^S_{ij}-2F^S_{ij}
\langle{\bf S}_i \cdot {\bf S}_j\rangle\right)
T^z_i T^z_j - \left(\frac{1}{2}F^D_{ij}+2F^D_{ij}
\langle{\bf S}_i \cdot {\bf S}_j\rangle\right)
{\bf T}^{\perp}_i \cdot {\bf T}^{\perp}_j\right].
\end{eqnarray}
To simplify the following equations we define effective couplings for the spin
and pseudospin Hamiltonians:
\begin{eqnarray}
\label{43}
J^S_{ij}&=&\frac{1}{2}F^S_{ij}+
2F^S_{ij}\langle T^z_i T^z_j\rangle +2F^D_{ij}\langle {\bf 
T}^{\perp}_i \cdot {\bf T}^{\perp}_j\rangle\nonumber\\
J^{Pz}_{ij}&=&2H_{ij}-\frac{1}{2}F^S_{ij}-2F^S_{ij}
\langle{\bf S}_i \cdot {\bf S}_j\rangle\nonumber\\
J^{P\perp}_{ij}&=&\frac{1}{2}F^D_{ij}+2F^D_{ij}
\langle{\bf S}_i \cdot {\bf S}_j\rangle.
\end{eqnarray}
In terms of these couplings (\ref{41}) and (\ref{42}) are written as
\begin{equation}
\label{44}
H^S=-\Delta_z\sum_iS^z_i-\sum_{ij}J^S_{ij}{\bf S}_i \cdot {\bf S}_j,
\end{equation}
\begin{equation}
\label{45}
H^P=-\sum_i\left[\Delta_V T^z_i+\Delta_t T^x_i\right]
+\sum_{ij}\left[J^{Pz}_{ij}T^z_i T^z_j - J^{P\perp}_{ij}
{\bf T}^{\perp}_i \cdot {\bf T}^{\perp}_j\right].
\end{equation}

We now analyze the effective spin and pseudospin Hamiltonians 
using the methods developed in section~\ref{s3}. 
Let us consider $H^S$ first.
As before, we represent spin operators by two Schwinger bosons, define 
ferromagnetic bond operators $F^{\dagger}_{ij}$
and rewrite the Hamiltonian in terms of these operators
\begin{equation}
\label{46}
H^S=-\frac{\Delta_z}{2}\sum_i\left(a^{\dagger}_{i1}a_{i1}-
a^{\dagger}_{i2}a_{i2}\right)-\frac{1}{2}\sum_{ij}J^S_{ij}
:F^{\dagger}_{ij}F_{ij}:.
\end{equation}
The main difference from the simple Heisenberg model we considered before 
is that here we have long-range spin-spin couplings.
Correspondingly, bond operators are also defined for all pairs of sites,
not just the nearest-neighbor ones.

We define expectation value of the bond operator for each pair of sites
\begin{equation}
\label{47}
Q^S_{ij}=\langle F^{\dagger}_{ij}\rangle = \langle F_{ij}\rangle,
\end{equation}
and perform a Hartree-Fock decoupling,  
adding a chemical potential term to account for the constraint
\begin{equation}
\label{48}
H^S=\sum_i\left[\left(\lambda-\frac{\Delta_z}{2}\right)a^{\dagger}_{i1}
a_{i1}+\left(\lambda+\frac{\Delta_z}{2}\right)a^{\dagger}_{i2}a_{i2}\right]-
\frac{1}{2}\sum_{ij}\tilde J^S_{ij}\left(a^{\dagger}_{i1}a_{j1}+
a^{\dagger}_{i2}a_{j2}+h.c.\right),
\end{equation}
where we have introduced notation $\tilde J^S_{ij}=J^S_{ij}Q^S_{ij}$.

Taking advantage of translational invariance we rewrite (\ref{48})
in Fourier space, simultaneously diagonalizing it and reducing 
the number of variables
\begin{equation}
\label{49}
H^S=\sum_{\bf k}\left[\left(\lambda-\frac{\Delta_z}{2}-
\tilde J^S_{\bf k}\right)a^{\dagger}_{{\bf k} 1}a_{{\bf k} 1}+
\left(\lambda+\frac{\Delta_z}{2}-
\tilde J^S_{\bf k}\right)a^{\dagger}_{{\bf k} 2}a_{{\bf k} 2}\right].
\end{equation}
As before, it is convenient to redefine $\lambda$ to make the notation 
more physical and also to facilitate numerical solution of the self-consistent
equations,
\begin{equation}
\label{50}
\lambda \rightarrow \lambda+\frac{\Delta_z}{2}+\tilde J^S_{\bf 0}. 
\end{equation}
Then $H^S$ becomes
\begin{equation}
\label{51}
H^S=\sum_{\bf k}\left(\epsilon_{{\bf k}1}a^{\dagger}_{{\bf k} 1}a_{{\bf k} 1}+
\epsilon_{{\bf k}2}a^{\dagger}_{{\bf k} 2}a_{{\bf k} 2}\right),
\end{equation}
where
\begin{eqnarray}
\label{52}
\epsilon_{{\bf k}1}&=&\lambda+\tilde J^S_{\bf 0}-\tilde J^S_{\bf k}\nonumber\\
\epsilon_{{\bf k}2}&=&\lambda+\Delta_z+\tilde J^S_{\bf 0}-\tilde J^S_{\bf k}.
\end{eqnarray}     
The two kinds of Schwinger bosons now have different dispersions due to the
asymmetry introduced by the magnetic field.
The $a^{\dagger}_2$ Schwinger bosons become the usual magnons in the limit 
$T\rightarrow 0$ and $N\rightarrow \infty$.
On the other hand, the bosons created by $a^{\dagger}_1$ Bose condense
in the same limit which signals the appearance of long-range order at $T=0$
in the thermodynamic limit.   
The connection between the condensation of Schwinger bosons and long-range 
magnetic order is discussed in more detail in.~\cite{Hirsch89}
We will always consider finite temperatures and finite system sizes
and therefore will not be concerned with this.

The system of self-consistent equations one has to solve now is as follows
\begin{eqnarray}
\label{53}
&&\frac{1}{N}\sum_{\bf k}\left[n_B(\epsilon_{{\bf k}1})+
n_B(\epsilon_{{\bf k}2})\right]=1\nonumber\\
&&Q^S_{\bf k}=n_B(\epsilon_{{\bf k}1})+n_B(\epsilon_{{\bf k}2}).
\end{eqnarray}
One therefore has $N+1$ coupled equations to solve instead of 2 in the case
of the Heisenberg model with nearest neighbor interactions.
It is useful to define another expectation value which will be used later,
\begin{equation}
\label{53.1}
\tilde Q^S_{\bf k}=n_B(\epsilon_{{\bf k}1})-n_B(\epsilon_{{\bf k}2}).
\end{equation}
These equations have to be solved self-consistently with the corresponding  
equations for the effective pseudospin Hamiltonian that differ from our 
detailed discussion of the near-neighbor interaction model in a similar manner.
The effective pseudospin Hamiltonian is
\begin{equation}
\label{54}
H^P=-\sum_i\Delta_t T^x_i
+\sum_{ij}\left[J^{Pz}_{ij}T^z_i T^z_j - J^{P\perp}_{ij}
{\bf T}^{\perp}_i \cdot {\bf T}^{\perp}_j\right],
\end{equation}
or, in terms of Schwinger bosons and bond operators
\begin{equation}
\label{55}
H^P=-\frac{\Delta_t}{2}\sum_i\left(b^{\dagger}_{i1}b_{i2}+b^{\dagger}_{i2}
b_{i1}\right)-\frac{1}{4}\sum_{ij}\left[\left(J^{P\perp}_{ij}-J^{Pz}_{ij}
\right):F^{\dagger}_{ij}F_{ij}:+\left(J^{P\perp}_{ij}+J^{Pz}_{ij}\right)
X^{\dagger}_{ij}X_{ij}\right].
\end{equation}
Here we use a different notation for Schwinger boson creation and annihilation
operators to distinguish them from the analogous operators representing 
real spins.
 
\begin{figure}
\begin{center}
\epsfxsize=4in
\epsffile{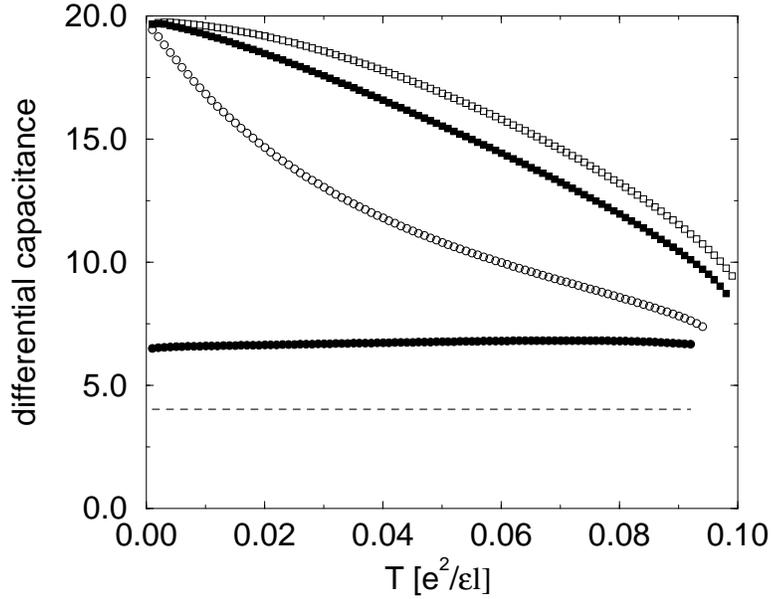}
\caption{Temperature dependence of the differential capacitance of 
the bilayer $d\langle T^z\rangle/d\Delta_V$ for $d/\ell=0.5$ and $\Delta_z=0.$ 
(filled circles),
$\Delta_z=0.001$ (open circles), $\Delta_z=0.005$ (filled squares) and 
$\Delta_z=0.01$ (open squares). 
The constant Hartree-Fock susceptibility is shown by a dashed line.
The large increase of the capacitance compared to the Hartree-Fock 
value is an artifact of SBMFT.
The main effect determining the temperature dependence of the capacitance 
at $d/\ell=0.5$ is
the suppression of spin-polarization by thermal fluctuations which influences 
effective pseudospin-pseudospin interactions (see Eq.(\ref{43})).} 
\label{difcapd0.5}
\end{center}
\end{figure}

\begin{figure}
\begin{center}
\epsfxsize=4in
\epsffile{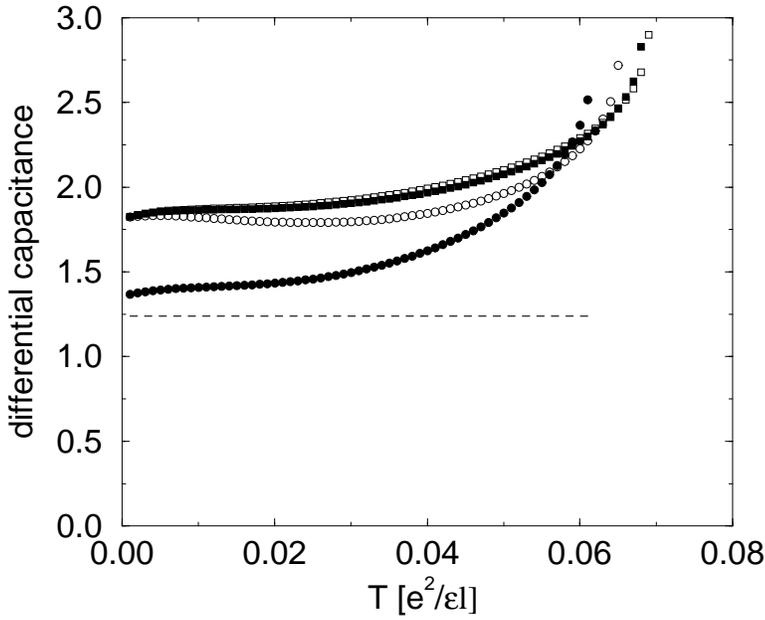}
\caption{Temperature dependence of the differential capacitance of 
the bilayer  
$d\langle T^z\rangle/d\Delta_V$ for $d/\ell=1.0$ and $\Delta_z=0.$ 
(filled circles),
$\Delta_z=0.001$ (open circles), $\Delta_z=0.005$ (filled squares) and 
$\Delta_z=0.01$ (open squares). 
The constant Hartree-Fock susceptibility is shown by a dashed line.
The influence of spin-polarization on the pseudospin system is weaker here 
compared to the case $d/\ell=0.5$ and the temperature dependence of the 
differential capacitance is mainly determined by the softening of the 
XY-ferromagnetic pseudospin order by thermal fluctuations, 
although there still is a substantial dependence on the Zeeman-coupling 
strength.}   
\label{difcapd1.0}
\end{center}
\end{figure}

\begin{figure}
\begin{center}
\epsfxsize=4in
\epsffile{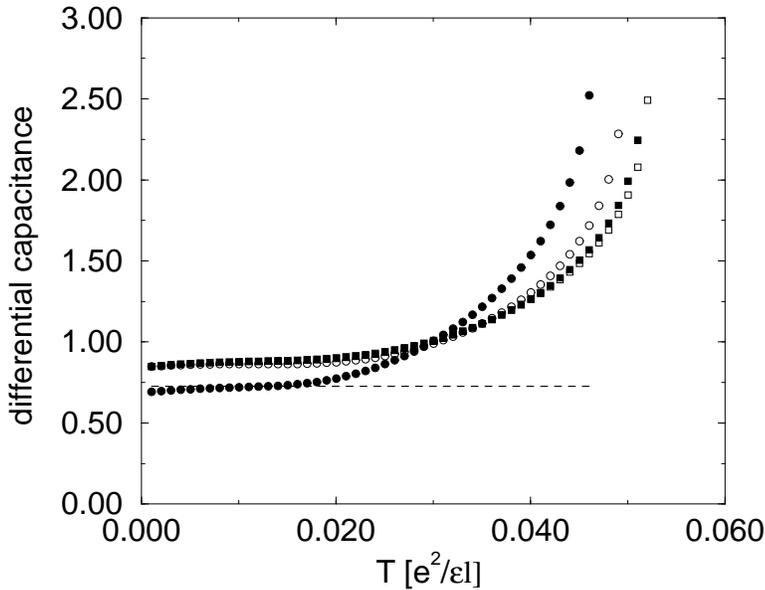}
\caption{Temperature dependence of the differential capacitance of 
the bilayer  
$d\langle T^z\rangle/d\Delta_V$ for $d/\ell=1.4$ and $\Delta_z=0.$ 
(filled circles),
$\Delta_z=0.001$ (open circles), $\Delta_z=0.005$ (filled squares) and 
$\Delta_z=0.01$ (open squares). 
The constant Hartree-Fock susceptibility is shown by a dashed line.
The spin-polarization influence has become very small both in the temperature
and Zeeman-coupling dependence.}   
\label{difcapd1.4}
\end{center}
\end{figure}

\begin{figure}
\begin{center}
\epsfxsize=4in
\epsffile{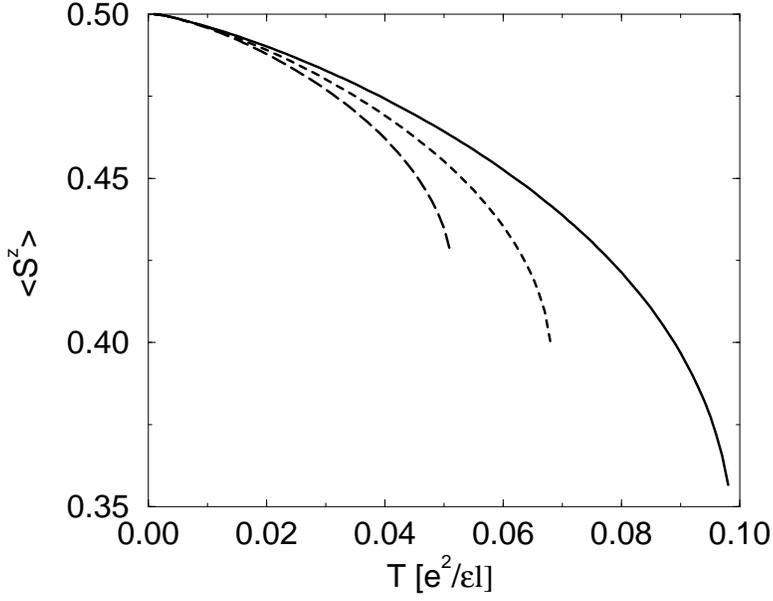}
\caption{Temperature dependence of the spin magnetization of the 
bilayer system 
for $\Delta_z=0.005$, $\Delta_t=0.001$ and $d/\ell=0.5$ (solid line), 
$d/\ell=1.0$ (dashed line)
and $d/\ell=1.4$ (long dashed line). The strong dependence of the spin 
magnetization on the interlayer separation is a signature 
of the spin-pseudospin coupling.}  
\label{spinmag}
\end{center}
\end{figure}

\begin{figure}
\begin{center}
\epsfxsize=4in
\epsffile{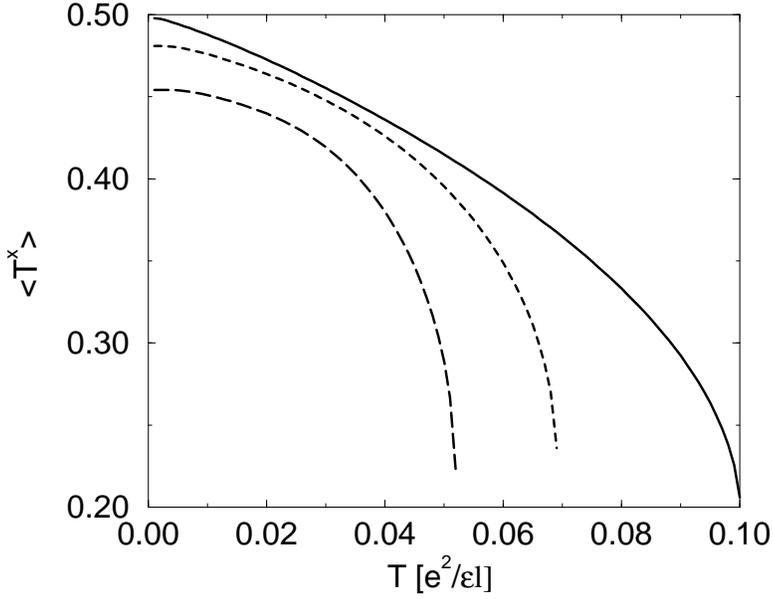}
\caption{Temperature dependence of the in-plane pseudospin magnetization 
of the bilayer for 
$\Delta_t=0.001$, $\Delta_z=0.005$ and $d/\ell=0.5$ (solid line), 
$d/\ell=1.0$ (dashed line) and $d/\ell=1.4$ (long dashed line). Pseudospin 
XY-ferromagnetic order is weakened by quantum fluctuations as the interlayer 
separation is increased.}  
\label{psmag}
\end{center}
\end{figure}

It is convenient to first rotate coordinates to diagonalize the tunneling 
term
\begin{equation}
\label{56}
H^P=-\frac{\Delta_t}{2}\sum_i\left(b^{\dagger}_{i1}b_{i1}-b^{\dagger}_{i2}
b_{i2}\right)-\frac{1}{4}\sum_{ij}\left[\left(J^{P\perp}_{ij}-J^{Pz}_{ij}
\right):F^{\dagger}_{ij}F_{ij}:+\left(J^{P\perp}_{ij}+J^{Pz}_{ij}\right)
X^{\dagger}_{ij}X_{ij}\right],
\end{equation} 
so that the bond operators are
\begin{eqnarray}
\label{57}
F^{\dagger}_{ij}&=&b^{\dagger}_{i1}b_{j1}+b^{\dagger}_{i2}b_{j2} \nonumber\\
X^{\dagger}_{ij}&=&b^{\dagger}_{i1}b^{\dagger}_{j1}-b^{\dagger}_{i2}
b^{\dagger}_{j2}.
\end{eqnarray}
Introducing expectation values for the bond operators,
\begin{eqnarray}
\label{58}
Q^P_{ij}&=&\langle F^{\dagger}_{ij}\rangle=\langle F_{ij}\rangle \nonumber \\
P^P_{ij}&=&\langle X^{\dagger}_{ij}\rangle=\langle X_{ij}\rangle,
\end{eqnarray}
then performing Hartree-Fock decoupling and introducing renormalized 
couplings, 
\begin{eqnarray}
\label{59}
\tilde J^-_{ij}&=&\frac{1}{2}\left(J^{P\perp}_{ij}-J^{Pz}_{ij}\right)Q^P_{ij},
\nonumber\\
\tilde J^+_{ij}&=&\frac{1}{2}\left(J^{P\perp}_{ij}+J^{Pz}_{ij}\right)P^P_{ij},
\end{eqnarray}
we obtain the following mean field Schwinger boson Hamiltonian
\begin{eqnarray}
\label{60}
H^P&=&\sum_i\left[\left(\lambda-\frac{\Delta_t}{2}\right)b^{\dagger}_{i1}
b_{i1}+\left(\lambda+\frac{\Delta_t}{2}\right)b^{\dagger}_{i2}b_{i2}\right]
\nonumber\\ 
&-&\frac{1}{2}\sum_{ij}\left[\tilde J^-_{ij}\left(b^{\dagger}_{i1}b_{j1}+
b^{\dagger}_{i2}b_{j2}+h.c.\right)+\tilde J^+_{ij}\left(b^{\dagger}_{i1}
b^{\dagger}_{j1}-b^{\dagger}_{i2}b^{\dagger}_{j2}+h.c.\right)\right].
\end{eqnarray}
Fourier transforming and redefining $\lambda$ as before 
\begin{eqnarray}
\label{61}
H^P&=&\sum_{\bf k}\left[\left(\lambda+\tilde J^+_{\bf 0}+\tilde J^-_{\bf 0}-
\tilde J^-_{\bf k}\right)b^{\dagger}_{{\bf k} 1}b_{{\bf k} 1}+
\left(\lambda+\Delta_t+\tilde J^+_{\bf 0}+\tilde J^-_{\bf 0}-
\tilde J^-_{\bf k}\right)b^{\dagger}_{{\bf k} 2}b_{{\bf k} 2}\right.\nonumber\\
&-&\left.\frac{1}{2}\tilde J^+_{\bf k}\left(b^{\dagger}_{{\bf k} 1}b^{\dagger}_
{{\bf -k} 1}-b^{\dagger}_{{\bf k} 2}b^{\dagger}_{{\bf -k} 2}+h.c.\right)\right].
\end{eqnarray}
As before, we introduce the notation
\begin{eqnarray}
\label{61.1}
A_{{\bf k}1}&=&\lambda+\tilde J^+_{\bf 0}+\tilde J^-_{\bf 0}-\tilde J^-_{\bf k}
\nonumber\\
A_{{\bf k}2}&=&\lambda+\Delta_t+\tilde J^+_{\bf 0}+\tilde J^-_{\bf 0}-
\tilde J^-_{\bf k}\nonumber\\
B_{{\bf k}1,2}&=&\pm\tilde J^+_{\bf k}\nonumber\\
\epsilon_{{\bf k}1,2}&=&\sqrt{A_{{\bf k}1,2}^2-B_{{\bf k}1,2}^2}.
\end{eqnarray}
Diagonalizing (\ref{61}) by Bogoliubov transformation we finally obtain
\begin{equation}
\label{62}
H^P=\sum_{\bf k}\left(\epsilon_{{\bf k}1}\beta^{\dagger}_{{\bf k}1}
\beta_{{\bf k}1}+\epsilon_{{\bf k}2}\beta^{\dagger}_{{\bf k}2}
\beta_{{\bf k}2}\right).
\end{equation}
As usual the parameters in the mean field Hamiltonain are fixed by solving the 
self-consistency equations
\begin{eqnarray}
\label{64}
&&\frac{1}{N}\sum_{\bf k}\left\{\frac{A_{{\bf k}1}}{\epsilon_{{\bf k}1}}
\left[n_B(\epsilon_{{\bf k}1})+\frac{1}{2}\right]+
\frac{A_{{\bf k}2}}{\epsilon_{{\bf k}2}}
\left[n_B(\epsilon_{{\bf k}2})+\frac{1}{2}\right]-1\right\}=1
\nonumber\\
&&Q^P_{\bf k}=\frac{A_{{\bf k}1}}{\epsilon_{{\bf k}1}}
\left[n_B(\epsilon_{{\bf k}1})+\frac{1}{2}\right]+
\frac{A_{{\bf k}2}}{\epsilon_{{\bf k}2}}
\left[n_B(\epsilon_{{\bf k}2})+\frac{1}{2}\right]-1
\nonumber\\
&&P^P_{\bf k}=\frac{B_{{\bf k}1}}{\epsilon_{{\bf k}1}}
\left[n_B(\epsilon_{{\bf k}1})+\frac{1}{2}\right]-
\frac{B_{{\bf k}2}}{\epsilon_{{\bf k}2}}
\left[n_B(\epsilon_{{\bf k}2})+\frac{1}{2}\right].
\end{eqnarray}
The systems of equations (\ref{53}) and (\ref{64}) have to be supplemented
by the equations for the spin-spin and pseudospin-pseudospin 
correlation functions, which couple the spin and pseudospin systems.
They can be evaluated straightforwardly using Wick's theorem.  For unbiased 
bilayers 
\begin{eqnarray}
\label{65}
&&\langle {\bf S}_i \cdot {\bf S}_j\rangle=\frac{3}{8}|Q^S_{ij}|^2-
\frac{1}{8}|\tilde Q^S_{ij}|^2+\langle S^z_i\rangle^2
\nonumber\\
&&\langle T^z_i T^z_j \rangle=\frac{1}{8}\left(|Q^P_{ij}|^2-|P^P_{ij}|^2-
|\tilde Q^P_{ij}|^2+|\tilde P^P_{ij}|^2\right)
\nonumber\\
&&\langle {\bf T}^{\perp}_i \cdot {\bf T}^{\perp}_j\rangle = \frac{1}{4}
\left(|Q^P_{ij}|^2+|P^P_{ij}|^2\right)+\langle T^x_i\rangle^2,
\end{eqnarray}
where
\begin{eqnarray}
\label{65.1}
\tilde Q^P_{\bf k}&=&\langle b^{\dagger}_{{\bf k}1}b_{{\bf k}2}+
b^{\dagger}_{{\bf k}2}b_{{\bf k}1}\rangle=
\frac{A_{{\bf k}1}}{\epsilon_{{\bf k}1}}
\left[n_B(\epsilon_{{\bf k}1})+\frac{1}{2}\right]-
\frac{A_{{\bf k}2}}{\epsilon_{{\bf k}2}}
\left[n_B(\epsilon_{{\bf k}2})+\frac{1}{2}\right]\nonumber \\
\tilde P^P_{\bf k}&=&\langle b_{{\bf k}1}b_{{\bf -k}1}+b_{{\bf k}2}
b_{{\bf -k}2}\rangle=
\frac{B_{{\bf k}1}}{\epsilon_{{\bf k}1}}
\left[n_B(\epsilon_{{\bf k}1})+\frac{1}{2}\right]+
\frac{B_{{\bf k}2}}{\epsilon_{{\bf k}2}}
\left[n_B(\epsilon_{{\bf k}2})+\frac{1}{2}\right].
\end{eqnarray}
Equations (\ref{53}), (\ref{64}), (\ref{65}) constitute the full set 
of equations of the SBMFT for the $\nu=1$ quantum Hall bilayer.
Fig.'s \ref{difcapd0.5}--\ref{psmag} summarize some of 
the results we have obtained by solving these equations numerically. 
The calculations were performed for lattices with up to 20$\times$20 sites. 

The polarization response of the pseudospin to a field that is 
transverse to the ordering direction corresponds
physically to the charge transferred between layers in response to a bias 
voltage.  In the Hartree-Fock approximation this quantity has no temperature 
dependence for temperatures below the charge gap.  
Our results for the temperature dependence of the differential capacitance 
at $d/\ell=0.5$, $1.0$ and $1.4$ at several different values
of the Zeeman coupling strength $\Delta_z$ are summarized in 
Fig.'s~\ref{difcapd0.5}--\ref{difcapd1.4}.
The limit $\Delta_z \to \infty$ corresponds to the usual assumption
that the spin degree of freedom is frozen out.   
The large increase of susceptibility compared to the Hartree-Fock
value at zero temperature in Fig.\ref{difcapd0.5} is an artifact of SBMFT,
which overestimates the value of the spin-spin correlation function in 
Eq.(\ref{43}) leading to an overestimate of the transverse susceptibility.
We did not try to fix this problem by normalizing the value of the 
correlation function by hand since we are mainly concerned with the 
qualitative features of the temperature dependence of susceptibility 
rather than exact values for its magnitude. 
     
One can see that the temperature dependences in Fig.\ref{difcapd0.5} are 
qualitatively different from those in the XY-model case in Fig.\ref{xysusc}.
The source of this difference is the dependence of the effective pseudospin 
interactions in Eq.(\ref{43}) on the spin polarization, as is 
evident from the strong dependence of the differential capacitance 
on Zeeman coupling.  It is clear from thermodynamic considerations that 
the transverse pseudospin coupling is approximately inversely 
proportional to the difference between the in-plane and transverse effective 
couplings.
From Eq.(\ref{43}) we see that both become less ferromagnetic 
(or more antiferromagnetic)
as correlations in the spin system are suppressed and the exchange 
contributions to the 
pseudospin effective interactions are reduced.   The coupling between 
spin and pseudospin fluctuations therefore tends to make the easy-plane anisotropy strengthen
and the transverse susceptibility weaken with increasing temperature. 
This effect competes with the increase in susceptibility with temperature that 
holds for the XY-model with temperature independent coupling constants.  
Our calculations indicate that the former effect dominates as small 
$d/\ell$ while
the latter effect dominates at larger $d/\ell$.  This qualitative 
discussion does not account for subtle effects associated with the longer range 
of the interaction between pseudospin $\hat z$ components, or for role 
played by the relative sign of the XY and Ising pseudospin interaction constants.
The transverse pseudospin susceptibility can be measured directly by measuring 
the capacitance of the bilayer system.~\cite{difcapref} 
We emphasize that the temperature dependences shown in these figures are all
due to correlated quantum and thermal fluctuations and would be absent in a 
Hartree-Fock theory.

In Fig.\ref{spinmag} and \ref{psmag} the temperature dependences of the spin 
and in-plane pseudospin magnetization for different values of interlayer 
separation are plotted.  
The spin density in bilayer systems can be measured by 
measuring the Knight shift in an optically pumped NMR experiment~\cite{nmrrefs}
while the pseudospin polarization can be extracted from optical absorption 
experiments.~\cite{psmagrefs}
The strong dependence of spin magnetization on the interlayer separation that 
we find is another signature of spin-pseudospin coupling.

\section{Concluding Remarks}
\label{s5}

In this paper we have presented a theory of the finite temperature properties
of $\nu=1$ quantum Hall bilayers that accounts for both spin and layer degrees 
of freedom.  The theory is based on an approximate mapping of the microscopic interacting electron 
problem onto a problem of interacting spins and pseudospins on a lattice
with a prescribed Hamiltonian specified in Eq.(\ref{14}).
Our mapping is exact when fluctuations in the charge density, summed over the 
two layers,
are neglected.  To investigate the finite temperature properties of this model,
we first generalize the Schwinger boson mean field theory of Arovas and Auerbach to anisotropic 
spin Hamiltonians.  Using this approach, we are able to 
predict qualitatively the character of the temperature dependence 
of several experimentally observable quantities like the spin magnetization 
and transverse pseudospin susceptibility, or interlayer differential 
capacitance.  The temperature dependences we find reflect a subtle interplay between
spin and pseudospin fluctuations.  Spin fluctuations in bilayer quantum Hall systems 
can be neglected only at temperatures significantly smaller than the unenhanced 
spin-splitting gap, which is $\sim 1K$ in typical circumstances. 
When spin fluctuations are neglected our calculation is closely related to
the microscopic generalized random-phase-approximation (RPA) calculation of 
Joglekar {\it et al}.~\cite{Yogesh01}
The linear spin-wave theory approximation to our spin-pseudospin model is equivalent to the 
microscopic generalized RPA; by neglecting charge fluctuations we are able to 
go beyond this level, approximating the spin-wave interactions of bilayer quantum
Hall systems using Schwinger boson mean field theory.  
Our spin-pseudospin model
is not however coupled in any way to any particular technique and the full array
of numerical and analytic methods that have been developed for quantum spin
models could be employed. 

Our formalism could in principle be generalized to integer filling factors greater than one.
To illustrate the complications involved we discuss briefly the case 
of $\nu=2$.
Quantum Hall bilayers at $\nu=2$ have been studied actively in recent years 
and a number of interesting theoretical results concerning broken symmetry 
ground states have also been 
obtained~\cite{Zheng97,Dassarma98,Dassarma99,Macdonald99,Schliemann00} for 
this case. 
In particular, a {\it canted antiferromagnetic} phase is predicted 
in which total spins in the two layers have opposite tilts away from the magnetic field
{\em and} spin-dependent spontaneous coherence is established between the opposite layers.
Thus spin-pseudospin interplay is important at $\nu=2$ even for ground state 
properties. 
At $\nu=2$ the single Slater determinant wavefunctions that span the Hilbert
space of the system have the following form
\begin{equation}
\label{66}
|\Psi[z]\rangle = \prod_{i,\alpha=1,2} \left(\sum_{k=1}^4 z_{ik}^{\alpha} 
c^{\dagger}_{ik}
\right)| 0 \rangle,
\end{equation}     
i.e. there are two occupied orbitals corresponding to each magnetic Wannier 
state $i$.   In this case, orthonormality conditions for these 
four orbitals provide four constraints on what will become coherent state
labels in the boson path integral.
Following this line, the $\nu=2$ spin-pseudospin Hamiltonian is 
\begin{eqnarray}
\label{67}
H&=&-\sum_{i,\alpha}\left[\Delta_V T^z_{i\alpha}+\Delta_t T^x_{i\alpha}+
\Delta_z S^z_{i\alpha}\right]
\nonumber\\
&+&\sum_{ij,\alpha\beta} \left[ (2H_{ij}-\frac{1}{2}F^S_{ij})T^z_{i\alpha}
T^z_{j\beta}-
\frac{1}{2}F^D_{ij}{\bf T}^{\perp}_{i\alpha} \cdot {\bf T}^{\perp}_{j\beta}-
\frac{1}{2}F^S_{ij}{\bf S}_{i\alpha} \cdot {\bf S}_{j\beta}\right.\nonumber\\
&-&\left.2 F^S_{ij}({\bf S}_{i\alpha} \cdot {\bf S}_{j\beta}) T^z_{i\alpha} 
T^z_{j\beta}-
2F^D_{ij}({\bf S}_{i\alpha} \cdot {\bf S}_{j\beta})({\bf T}^{\perp}_{i\alpha} 
\cdot {\bf T}^{\perp}_{j\beta})\right].
\end{eqnarray} 
Each lattice site now has two spins and two pseudospins corresponding 
to two occupied orbitals in (\ref{66}). 
It would be interesting to study the correspondence between this 
Hamiltonian and two similar effective models introduced 
recently.~\cite{Demler99,Yang99}
Eq.(\ref{67}) is substantially more complicated than 
(\ref{14}) and 
even the classical ground state, again identical to the Hartree-Fock 
ground state, is known only numerically for general external fields.
The presence of implicit orthonormality constraints in Eq.(\ref{67}) 
is a disadvantage of this approach, since it reduces the transparency
of the Hamiltonian.  We believe, nevertheless,  that Eq.(\ref{67}) could be
be a good starting point for a theory of correlations and thermal 
fluctuations in $\nu=2$ quantum Hall ferromagnets.  

\section*{Acknowledgments}
We are grateful to Yogesh Joglekar and John Schliemann for many useful
discussions. 
This work was supported by the Welch Foundation, by the Indiana 21st Century 
Fund, and by the National Science Foundation under grant DMR0115947.

\section*{Appendix}
In this section we explain in more detail how our effective interactions are 
evaluated.
A key step is the contruction of the magnetic Wannier basis, and we start by 
briefly summarizing the results of Rashba {\it et al}.~\cite{Efros}
We want to construct a complete set of Wannier functions localized 
at the sites of a square lattice with lattice constant $a=\sqrt{2 \pi \ell^2}$.
We start from an overcomplete set of minimum uncertainty wave packets 
centered on square lattice sites. 
The minimum uncertainty wavepacket localized at the origin is 
\begin{equation}
\label{68}
c_{00}({\bf r}) = \frac{1}{\sqrt{2 \pi \ell^2}} 
\exp\left(-\frac{r^2}{4 \ell^2}\right).
\end{equation}
$c_{00}$ is just the zero angular momentum eigenfunction in the symmetric 
gauge.
To construct wavepackets localized at other sites of the square lattice we
need to translate $c_{00}$ wavepacket from the origin to each of the lattice
sites. 
Let the lattice sites be
\begin{equation}
\label{69}
{\bf r}_{mn}\,=\,m a \hat x + n a \hat y.
\end{equation}
Define
\begin{equation}
\label{70}
c_{mn}({\bf r}) = T_{ma\hat x}T_{na\hat y} c_{00}({\bf r}).
\end{equation}
Here 
\begin{equation}
\label{71}
T_{\bf R} = e^{-\frac{i}{\hbar}{\bf R}\left({\bf p}-\frac{e}{c}{\bf A}
\right)}
\end{equation}
is the magnetic translation operator (this expression is valid in the 
symmetric gauge only).
In the symmetric gauge
\begin{equation}
\label{72}
{\bf A} = \frac{B}{2}(-y,x,0).
\end{equation}
Substituting (\ref{72}) into (\ref{71}) one obtains
\begin{equation}
\label{73}
T_{ma\hat x}T_{na\hat y} = (-1)^{mn}\exp \left[\frac{i}{2\ell^2} \hat z 
\cdot ({\bf r} \times {\bf r}_{mn})\right] \exp\left(-\frac{i}{\hbar}
{\bf r}_{mn} \cdot {\bf p}\right).
\end{equation}
Therefore $c_{mn}$ is given by
\begin{equation}
\label{74}
c_{mn}({\bf r}) = \frac{(-1)^{mn}}{\sqrt{2\pi \ell^2}}
\exp\left[ -\frac{({\bf r} -{\bf r}_{mn})^2}{4 \ell^2} +
\frac{i}{2\ell^2} \hat z ({\bf r} \times {\bf r}_{mn})\right].
\end{equation}
The functions $c_{mn}$ are not orthogonal and form an overcomplete set
due to the following identity established by Perelomov
\begin{equation}
\label{75}
\sum_{mn=-\infty}^{\infty} (-1)^{m+n} c_{mn}({\bf r})=0.
\end{equation}

First we will construct a complete set of orthonormal Bloch functions 
from the set $\{c_{mn}\}$.
Assume a normalization plaquette of area $L^2=2\pi \ell^2 N_{\phi}$,
where $N_{\phi}$ is the number of magnetic flux quanta.
Define a Bloch function at quasimomentum ${\bf k}$ as follows
\begin{equation}
\label{76}
\Psi_{{\bf k}}({\bf r})=\frac{1}{\sqrt{N_{\phi}\nu({\bf k})}}
\sum_{mn=-\infty}^{\infty} c_{mn}({\bf r}) \exp (i {\bf k} {\bf r}_{mn}).
\end{equation}
Here $\nu({\bf k})$ is a momentum-dependent normalization factor.
The allowed values of quasimomentum are determined from the boundary
conditions. 
If we assume periodic boundary conditions with respect to magnetic 
translations $T_{ma\hat x}, T_{na\hat y}$ the allowed values 
are $k_{x,y}=2 \pi n_{x,y}/L$.

The Bloch functions are assumed to be normalized to unity over the 
normalization plaquette. 
\begin{equation}
\label{77}
\int_A d{\bf r} \left| \Psi_{{\bf k}}({\bf r}) \right |^2 = 1.
\end{equation}
The normalization factor is given by
\begin{equation}
\label{79}
\nu({\bf k})=a \sum_{mn=-\infty}^{\infty} c_{mn}(0)\cos({\bf k}{\bf r}_{mn}).
\end{equation} 
It turns out that $\nu({\bf k})$ goes to zero at the corners of the Brillouin
zone $k_{x,y}=\pm \pi/a$.
At these points the Bloch function has to be calculated by a careful limiting 
procedure (for details see~\cite{Efros}). 
The result is
\begin{equation}
\label{86}
\Psi_{{\bf k}_0}({\bf r}) = \frac{i}{a \sqrt{2 N_{\phi} \gamma}}
\sum_{mn=-\infty}^{\infty} (-1)^{m+n} \overline z_{mn} c_{mn}({\bf r}),
\end{equation}
where $\gamma = -\frac{1}{a} \sum_{mn=-\infty}^{\infty} (-1)^{m+n}
c_{mn}(0) x_m^2$. 
Now we can construct magnetic Wannier functions by inverse Fourier transform
\begin{equation}
\label{87}
W_{mn}({\bf r})=\frac{1}{\sqrt{N_{\phi}}} \sum_{{\bf k}} \Psi_{{\bf k}}
({\bf r})
\exp(-i{\bf k} {\bf r}_{mn}).
\end{equation} 
The functions $W_{nm}$ are orthonormal and form a complete set by construction.
We evaluated our effective spin and pseudospin interactions numerically by 
inserting these functions in direct and exchange Coulomb matrix elements
$\langle ij|V_{S,D}|ij\rangle$ and $\langle ij|V_{S,D}|ji\rangle$,
where $i=(nm)$. 
The calculation is best done by Fourier transforming the Coulomb interactions
$V_{S,D}$
\begin{eqnarray}
\label{88}
V_S({\bf q})\,&=&\,\frac{2\pi e^2}{\epsilon q} \nonumber \\
V_D({\bf q})\,&=&\,\frac{2\pi e^2}{\epsilon q}e^{-qd},
\end{eqnarray}
and evaluating the plane-wave matrix elements 
$\langle i|e^{i{\bf q r}}|j\rangle$.
The Coulomb interaction matrix elements are then given by
\begin{equation}
\label{89}
\langle ij|V_{S,D}|ij\rangle=\frac{1}{L^2}\sum_{\bf q}\langle i \left| 
e^{i{\bf q r}} \right|
i \rangle\, V_{S,D}({\bf q})\, \langle j \left| e^{-i{\bf q r}} \right| j 
\rangle,
\end{equation}
and 
\begin{equation}
\label{90}
\langle ij|V_{S,D}|ji\rangle=\frac{1}{L^2}\sum_{\bf q}\langle i \left| 
e^{i{\bf q r}} \right|
j \rangle\, V_{S,D}({\bf q})\, \langle j \left| e^{-i{\bf q r}} \right| i 
\rangle.
\end{equation}

\newpage

\end{document}